\documentclass[]{jfm}

\usepackage[format = plain, singlelinecheck = no, justification = justified]{caption}
\usepackage{graphicx}
\usepackage{newtxtext}
\usepackage{newtxmath}
\usepackage{natbib}
\usepackage{hyperref}
\usepackage[dvipsnames]{xcolor}
\usepackage[normalem]{ulem}

\usepackage[inkscapelatex=false]{svg}

\usepackage{dcolumn}
\usepackage{bm}
\usepackage{caption}
\usepackage[symbol]{footmisc}

\hypersetup{
    colorlinks = true,
    urlcolor   = blue,
    citecolor  = black,
}

\newcommand{\RomanNumeralCaps}[1]

\title{Reference map technique for Lagrangian  exploration of coherent structures}

\author{Imran Hayat\aff{1}\thanks{These two authors are the co-first  
 authors who contributed equally to this work.}
  Ryan T. Black\aff{1}\footnotemark[1]
 \and George Ilhwan Park\aff{1}
 \corresp{\email{gipark@seas.upenn.edu}},
 }

\affiliation{\aff{1}Department of Mechanical Engineering and Applied Mechanics, University of Pennsylvania, Philadelphia, PA 19104, USA\\}

\begin{document}
\maketitle

\begin{abstract}

We explore the application of the reference map technique, originally developed for Eulerian simulation of solid mechanics, in Lagrangian kinematics of fluid flows. Unlike traditional methods based on explicit particle tracking, the reference map facilitates calculation of flow maps and gradients without the need for particles. This is achieved through an Eulerian update of the reference map, which records the take-off positions of fluid particles. Using the reference map, measures of Lagrangian fluid-element deformation can be computed solely from Eulerian data, such as displacement, deformation gradient, Cauchy-Green tensor, and Finite-Time Lyapunov Exponent (FTLE) fields. We first demonstrate the accuracy of FTLE calculations based on the reference map against the standard particle-based approach in a 2D Taylor-Green vortex. Then, we apply it to turbulent channel flow at $Re_\tau=180$ where Lagrangian coherent structures identified as ridges of the backward-time FTLE are found to bound vortical regions of flow,  consistent with Eulerian coherent structures from \color{black} the \color{black} Q-criterion. The reference map  also proves suitable for material surface tracking despite not explicitly tracking particles. This capability can provide valuable insights into the Lagrangian landscape of turbulent momentum transport, complementing Eulerian velocity-field analysis. The evolution of initially wall-normal material surfaces in the viscous sublayer, buffer layer, and log layer sheds light on the Reynolds stress-generating events from a Lagrangian perspective. Eliminating the need for tracking numerous particles, the reference map approach offers a convenient and promising avenue for future investigations into Lagrangian kinematics and dynamics of fluid flows.

\end{abstract}

\begin{keywords}
Authors should not enter keywords on the manuscript, as these must be chosen by the author during the online submission process and will then be added during the typesetting process (see \href{https://www.cambridge.org/core/journals/journal-of-fluid-mechanics/information/list-of-keywords}{Keyword PDF} for the full list).  Other classifications will be added at the same time.
\end{keywords}

{\bf MSC Codes }  {\it(Optional)} Please enter your MSC Codes here

\section{\label{sec:intro}Introduction}






Fluid flows are often characterized by dynamic behaviors of intricate spatial patterns. 
 Notable examples include bacteria-induced active-matter chaos \citep{ran2021bacteria, urzay2017multi}, a forest of vigorous turbulent eddies in wing boundary layers \citep{goc2021large}, and the great red spot of Jupiter \citep{marcus1993jupiter}. While visually captivating, the complex interaction among these spatial patterns can sometimes obscure straightforward comprehension of the dynamically important mechanisms governing them. The concept of coherent structures has proven integral to identifying these organized features and furthering the analysis and modeling of their dynamical impact on the flow. Coherent structures, broadly defined, represent persistent patterns or organized motions that stand out against the background flow. In a more specific sense, \cite{hussain_1986} defined coherent structures in turbulent flows as \color{black}``\color{black} a connected turbulent fluid mass with instantaneously phase-correlated vorticity over its spatial extent." Meanwhile, \cite{haller_lagrangian_2015} offered a definition strongly rooted in Lagrangian/material kinematics and dynamical systems theory, describing them as \color{black}``\color{black}special surfaces of fluid trajectories that organize the rest of the flow into ordered patterns." Despite differences in the perspectives and approaches taken to define them, it is perhaps unanimously agreed that coherent structures serve as the engines driving the complex interplay of momentum, heat, and scalar transport at various regimes.

Recognizing the crucial role coherent structures play in fluid dynamics, the fluids community has dedicated considerable attention to developing methods for elucidating and understanding these organized features. A widely used class of Eulerian methods, particularly within the turbulence community, involves vortex identification schemes based on the invariants of the velocity gradient tensor or its symmetric/antisymmetric decomposition.
The well-known \color{black} $Q$-criterion \color{black} \citep{hunt1988eddies} extracts vortical structures from the second invariant of the velocity gradient tensor $\nabla u$, defined as $Q = -\frac{1}{2} \frac{\partial u_i}{\partial x_j} \frac{\partial u_j}{\partial x_i} = \frac{1}{2}(||\Omega||^2 - ||S||^2)$, where $\Omega$ and $S$ are the rate-of-rotation and rate-of-strain tensors, respectively. $Q$ is regarded as the strength of the rotational component of the velocity field relative to the shear, and isosurfaces of $Q$ at a selected positive threshold value are used as candidate coherent structures.
In response to potential inaccuracies of \color{black} the $Q$-criterion \color{black} when vortices are subjected to strong external strain, \cite{jeong1995identification} proposed the $\lambda_2$-criterion. 
The $\lambda_2$-criterion imposes the condition that $\lambda_2 < 0$, where $\lambda_2$ is the median of three eigenvalues of $S^2 + \Omega^2$. This requirement is based on the principles of Galilean invariance and the presence of net vorticity, which is correlated with a local pressure minimum at the vortex core. 
\cite{zhou1999mechanisms} proposed a detection scheme based on 
complex eigenvalues of $\nabla u$.
 From the conjugate pair of the complex eigenvalues $\lambda_{cr} \pm i \lambda_{ci} $ of $\nabla u$, the imaginary part $\lambda_{ci}$ was shown to represent the local swirling strength of the vortex.
 In an extension of this work, \cite{chakraborty2005relationships} suggested an additional constraint $\lambda_{cr}/\lambda_{ci} < \delta$ to ensure the compactness of the spiral material orbits. 
The aforementioned  detection methods are computed from the 
instantaneous Eulerian velocity/tensor fields, making their application to spatial velocity data obtained from numerical simulations or experiments straightforward. They have proven effective in excluding regions of strong shear without net swirling motion. 
These methods necessitate users to prescribe arbitrary thresholds for isosurface detection. As demonstrated by \cite{pierce2013application}, all the aforementioned schemes produce nearly identical landscapes of turbulent eddies with an appropriate selection of threshold values for each scheme.

In contrast to Eulerian methods described in the previous paragraph, Lagrangian coherent structures (LCS) offer an alternative approach to coherent structure detection with an emphasis on coherence at the material (Lagrangian) level and objectivity. \cite{haller2005objective} demonstrated that the previously mentioned Eulerian diagnostics, while Galilean invariant, fail the test of objectivity, where the coherent structures identified should be invariant under Euclidean coordinate changes, including time-dependent frame rotation and translation. It's essential to note that this does not diminish the utility of Eulerian diagnostics altogether. These approaches effectively reveal coherent patterns in the instantaneous velocity field, aiding in understanding vortex interaction/formation and momentum transport in wall-bounded flows \citep{jeong1995identification, zhou1999mechanisms, duguet2012self, motoori2019generation, yao2020physical, chong1998turbulence}. However, limitations may arise in the presence of frame rotation, and structural models built from the detected patterns may have constraints concerning objectivity.
Haller recognized that objectivity can be ensured by basing coherent structure detection on fluid particle trajectories, which are fundamentally invariant Lagrangian features of a given flow. The representation of trajectories may vary in different reference frames, but the trajectories themselves remain invariant. The spatial patterns created by passively advected tracers then naturally embody the candidates for LCS.  \cite{haller_lagrangian_2015} explains LCS as \color{black}``\color{black}the most repelling, attracting, and shearing material surfaces that form the skeleton of Lagrangian particle dynamics." As such,  LCS is expected to have  direct connections to the motion of Lagrangian fluid elements. 

A widely used diagnostic tool for LCS is the finite-time Lyapunov exponent (FTLE) field. The FTLE characterizes the largest average stretching of an infinitesimal material line over a finite time interval. The extremizing surfaces of the FTLE field (commonly known as ridges) contain the candidates for LCS, which are indicators of locally most repelling or attracting material surfaces. 
The direction of time is important in this context; repelling LCS at a given time are computed from future data, while attracting LCSs require past data. Attracting LCSs often correspond to structures seen in flow visualization, such as near-wall turbulent eddies \citep{green2007} or von Karman vortex streets behind a cylinder \citep{kasten2010lagrangian}. 
However, FTLE ridges are  a necessary condition marking LCS candidates, and may fail to differentiate between attracting/repelling and shearing effects \citep{huang2022lagrangian}. 
Sufficient and necessary conditions for hyperbolic LCSs have been proposed in terms of invariants of the Cauchy–Green strain tensor field and variational algorithms \citep{haller_lagrangian_2011, farazmand2012computing, haller2012geodesic, farazmand2013attracting}. Although more rigorous and complete, calculating LCS satisfying the full conditions poses algorithmic and computational challenges (especially in 3D), and FTLE ridges continue to be used frequently for visual diagnostics of LCS.

The calculation of the FTLE field can be computationally expensive due to the necessity of accurately reconstructing a large number of particle trajectories. Typically, tracers, initially placed surrounding the fluid grid points, evolve by integrating in time their Lagrangian equation of motion. This integration is performed using the available Eulerian velocity data interpolated to the tracer locations. Subsequently, these trajectories are utilized to calculate the deformation gradient through finite difference methods, where the current tracer positions are differentiated with respect to their respective take-off positions. 
For two dimensional flows, 
\cite{ONU201526} suggests using more than 1.25 million tracers for a  fluid grid comprised of 0.25 million points.  
\cite{huang2022lagrangian} employed 80,000 tracer particles to investigate LCSs in the flow past a backward-facing step, 
where the two-dimensional flow simulation required 48,900 grid cells.
In the study by \cite{green2007} focusing on turbulent channel flow, the particle grid resolution was six times greater in all three dimensions compared to the DNS grid resolution. 

The application of LCS to three-dimensional turbulent flows is rare compared to a large volume of studies routinely applying Eulerian methods for coherent structure detection. \cite{green2007} and \cite{rockwood2018tracking} visualized a hairpin vortex in turbulent channel flow at $Re_{\tau} = 180$ and tracked individual coherent structure motion using Lagrangian saddle points. \cite{pan2009identification} investigated orientation angle and convection velocity of LCS from a time-resolved 2D particle image velocimetry (PIV) measurement of a turbulent boundary layer. \cite{bettencourt2013characterization} used the finite-size Lyapunov exponent (FSLE) to characterize LCSs in a turbulent channel flow. \cite{wilson_tutkun_cal_2013} highlighted LCSs of a flat-plate boundary layer (FPBL) at $Re_\theta = 9800$ at $y^+ = 50$ from stereo PIV measurements on a wall-parallel plane. \cite{he_pan_feng_gao_wang_2016} detailed evolution of LCSs in a FPBL transition induced by the wake of a circular cylinder using 2D PIV measurements. \cite{halic_krug_haller_holzner_2019} studied Lagrangian vortical coherent structures in a gravity current and their influence on the turbulent/non-turbulent interface and entrainment using three-dimensional particle tracking velocimetry (3-D PTV). \cite{thomas2022eulerian} studied spatio-temporal evolution of LCSs in a positive surge at Froude 1.5, using 2D stereo-PIV in a streamwise/free-surface-normal plane. All of the aforementioned studies used Lagrangian tracers as a basis for LCS computation. Half of them reported volumetric LCSs from full three-dimensional flow data \citep{green2007, rockwood2018tracking, bettencourt2013characterization, halic_krug_haller_holzner_2019}, whereas the rest focused on LCSs restricted to two-dimensional test planes. This might be due to the high cost of tracer trajectories calculations.

The purpose of this paper is to explore an alternative, particle-free approach for detecting Lagrangian coherent structures (LCS). The reference map technique, originally developed for Eulerian treatment of solid mechanics or fluid-structure interaction, will be adapted for fluid, and the steps for computing the Finite-Time Lyapunov Exponent (FTLE) field from the reference map will be delineated. Material line/surface tracking capabilities naturally embedded in the reference map will be discussed. The method will be first validated in a simple two-dimensional flow. An application to three-dimensional turbulence will be showcased in a low Reynolds number channel flow at $Re_\tau = 180$.




\section{\label{sec:RMT_theory}Theoretical background}

The discussion of the FTLE, LCS, and the reference map naturally necessitates tools from (Lagrangian) continuum mechanics. The use of these tools in the fluid mechanics community is currently not uniformly widespread. In this section, we begin  by reviewing definitions and concepts that are fundamental to describing the motion and deformation of continuum bodies from a Lagrangian perspective.

\begin{figure}
    \centering
    \captionsetup{width=\linewidth}
    \includegraphics[scale=0.5]{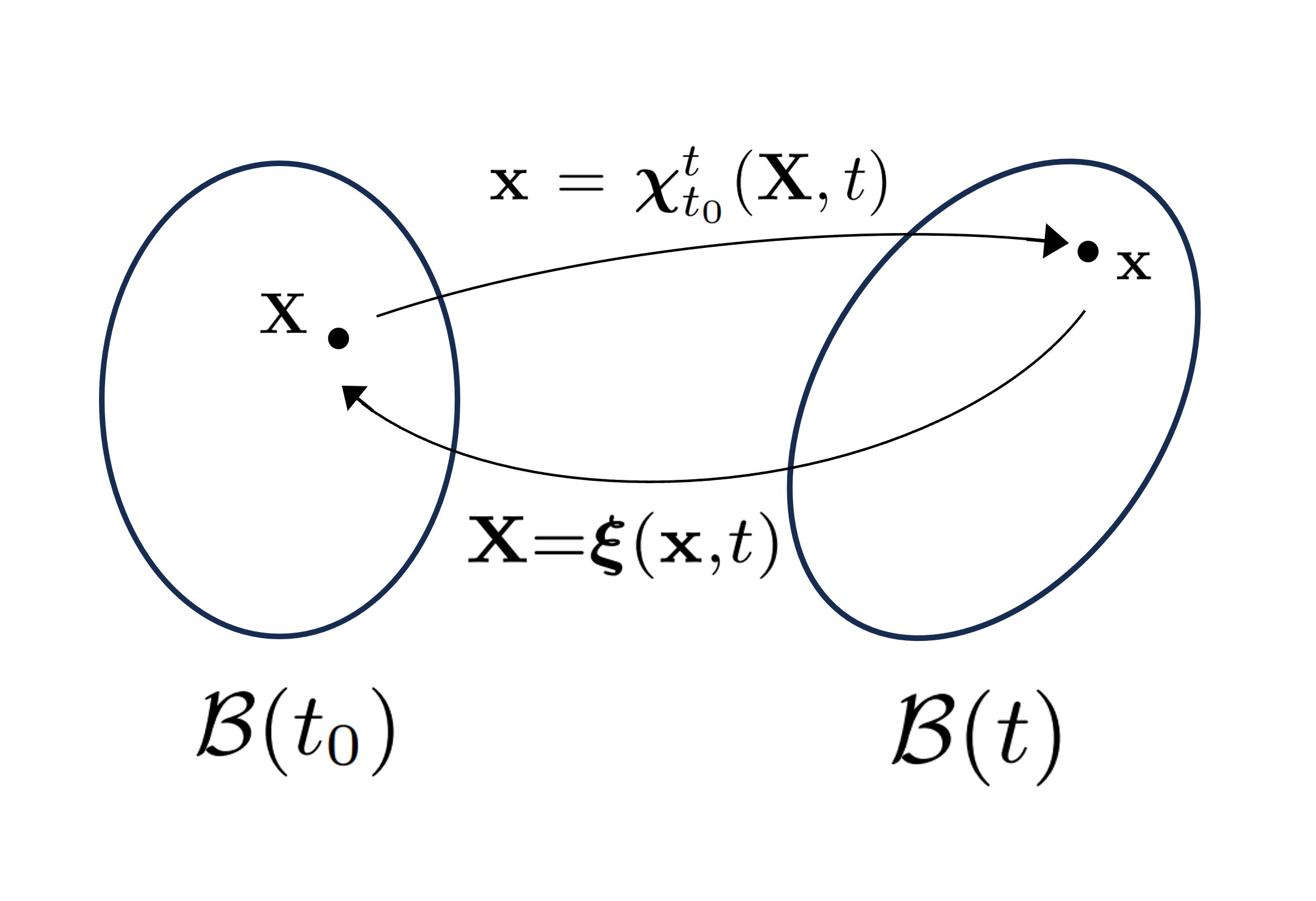}
    \caption{Motion map $\boldsymbol{\chi}^{t}_{t_0}(\textbf{X},t)$ from the reference configuration $\mathcal{B}(t_0)$ to the current configuration $\mathcal{B}(t)$,  and inverse motion map (or reference map) $\boldsymbol{\xi}(\textbf{x},t)$ from the current configuration to the reference configuration.}
    \label{rmt_map}
\end{figure}

First, the flow map (also referred to as deformation mapping or motion map) \color{black} \textbf{x} = \color{black}$\boldsymbol{\chi}^{t}_{t_0}(\textbf{X},t)$ maps a fluid particle (tracer) located at $\textbf{X}$ in the reference (or initial) configuration at the initial time $t_0$\color{black}, $\mathcal{B}(t_0)$, \color{black} to a point $\textbf{x}$ in the current configuration of the continuum body at time $t$\color{black}, $\mathcal{B}(t)$, \color{black}(see figure \ref{rmt_map}).
The deformation gradient (or Jacobian of the flow  map) $\mathbf{F}^{t}_{t_0}
(\textbf{X},t)$ is defined as 
\begin{equation}
    \mathbf{F}^{t}_{t_0}(\textbf{X},t) \equiv \frac{\partial \boldsymbol{\chi}^{t}_{t_0}(\textbf{X},t)}{\partial \textbf{X}}
    =
    \frac{\partial \textbf{x}}{\partial \textbf{X}}, 
\end{equation}
which is an important kinematic quantity for the Lagrangian description of motion and constitutive modeling of solid materials. 
For instance, $\bold{F}^{t}_{t_0}(\textbf{X},t)$ fully characterizes  changes in volume, area, and length of the infinitesimal material volume/area/line elements initially placed at $\textbf{X}$. Additionally, Cauchy stress for hyperelastic materials undergoing finite deformations is determined by the deformation gradient. 

A basic continuum mechanics hypothesis is that the flow map is one-to-one \citep{gurtin_mechanics_2010}. Thus, at each time $t$,  the flow map has an inverse, 
\begin{equation}
\boldsymbol{\xi}(\textbf{x},t) = \boldsymbol{\chi}^{t_0}_{t}(\textbf{x},t) = \textbf{X},     
\end{equation}
which maps a material point 
$\textbf{x}$ in the current configuration at time $t$ to its initial position $\textbf{X}$ in the reference configuration. 
Hereinafter, this inverse flow map $\boldsymbol{\xi}$ 
is referred to as the reference map following \color{black} \cite{gurtin_mechanics_2010} \color{black}. 
The deformation gradient $\textbf{F}^{t}_{t_0}(\textbf{X},t)$, 
which is the Jacobian of the forward motion map at the initial position $\textbf{X}$,
\begin{equation}
    \left. \textbf{F}^{t}_{t_0} (\textbf{X},t) \right\vert_{\textbf{X} = \boldsymbol{\xi}(\textbf{x},t)} = \left(\frac{\partial\boldsymbol{\xi}(\textbf{x},t)}{\partial \textbf{x}}\right)^{-1}, 
    \label{def_grad_rmt}
\end{equation}
 \color{black} can be determined from the reference map\color{black}. Analogously, the inverse deformation gradient (or the Jacobian of the backward motion map at the current position $\textbf{x}$) is defined as
\begin{equation}
    \textbf{F}^{t_0}_{t}(\textbf{x}) \equiv 
        \frac{\partial\boldsymbol{\xi}(\textbf{x},t)}{\partial \textbf{x}} = 
    \frac{\partial \textbf{X}}{\partial \textbf{x}}. 
    \label{grad_rmt}
\end{equation}
The reference map framework was developed originally in the context of isothermal compressible and incompressible fluid-structure interaction \citep{kamrin_reference_2012,valkov_eulerian_2015,jain_conservative_2019,rycroft_reference_2020, wang2022incompressible}. Additionally, it has been employed in computational solid mechanics with non-trivial hyperelastic constitutive models undergoing inhomogeneous deformation \citep{kamrin_reference_2012}. In this line of work, the Lagrangian solid equation recast in the Eulerian form, which resembles the Navier-Stokes equations, was solved using finite difference methods. The deformation gradient, determining the solid stress, was updated from the reference map.
The key starting point was to identify $\textbf{x}$, the current position of material points, as the fixed grid points in Eulerian simulations or measurements. $\boldsymbol{\xi}(\textbf{x},t)$ then returns the initial position $\textbf{X}$ of a Lagrangian material point that is arriving at the grid point $\textbf{x}$ at time $t$. Since a material point labeled by $\textbf{X}$ preserves its identity during the motion, the Lagrangian time derivative of the reference map is \color{black} the zero vector,\color{black}
\begin{equation}
    \frac{D\boldsymbol{\xi}}{D t} = 
    \frac{\partial\boldsymbol{\xi}}{\partial t} + \textbf{v}\cdot\nabla\boldsymbol{\xi} = \bm{0}, 
    \label{eq:rmt_update}
\end{equation}
where 
$\textbf{v} = \textbf{v}(\textbf{x},t)$ represents the spatial (Eulerian) velocity field, ensuring that the Lagrangian velocity of a material particle is equal to the Eulerian velocity taken at the instantaneous particle position. The gradient is taken with respect to $\textbf{x}$. Eulerian updates of the reference map (Eq.~(\ref{eq:rmt_update})) and conservation laws then enabled purely Eulerian simulations of solid mechanics and fluid-structure interaction using a fixed grid.

It is worth noting that there is no explicit notion of tracking particles with fixed identities forward in time in the reference map approach. This is in clear contrast to the particle-tracking-based method. Instead, the reference map approach assumes that particles (tracers) with different identities are arriving at the grid points at each time, and it endeavors to accurately remember where they took off from. In this sense, the reference map tracks particles backward in time but in an implicit manner.

In the present study, we employ the reference map to compute the Lagrangian kinematics of a fluid, where the reference map is updated through Eq.(\ref{eq:rmt_update}) in an Eulerian manner. That is, the reference map is passively advected on the same fixed Eulerian grid utilized in fluid dynamics simulations or from experimental velocity measurements. The deformation gradient is then determined entirely from the reference map, which can be further processed to compute Lagrangian coherent structures (LCSs). Additionally, the reference map provides a natural means for material surface tracking. Details of these two tasks are provided in the following subsections. Computational details associated with the discretization of Eq.(\ref{eq:rmt_update}) and initial/boundary conditions for the reference map are provided in Section \ref{sec:numerics}.

\color{black} Note that an Eulerian method to compute FTLE fields utilizing Eq.~(\ref{eq:rmt_update}) has been previously proposed by \cite{leung_eulerian_2011} in the context of the level set method, with application to simple flows \citep{leung2013backward, you2014eulerian, you2018eulerian, you2018improved, you2020fast, you2021eulerian}. 
Here, we connect this approach to the reference map technique,  and apply it to complex three-dimensional turbulent flows for the first time. Furthermore, we discretize the conservative form of Eq.~(\ref{eq:rmt_update}) utilizing low dissipation numerical methods important for scale-resolving simulation of  turbulent fluid flow  (i.e., a second order central  scheme), as compared to dissipative schemes  (e.g., WENO) used by previous studies. We also note that the use of the reference map for Lagrangian material surface tracking to investigate turbulent flow structures and momentum transport  in wall-bounded flows is a novelty that differs from previous applications of this technique. \color{black}

\subsection{\label{sec:theory_LCS}Computation of Lagrangian coherent structure (LCS)}
Identification of coherent structures is a crucial tool for assessing and characterizing fluid flows. In the Lagrangian Coherent Structures (LCS) approach, the flow field is partitioned based on material surfaces, allowing for the objective identification of ordered flow patterns. In this work, we adopt the definition of LCSs using the Finite-Time Lyapunov Exponent (FTLE) fields, which characterize the rate of stretching between two neighboring fluid particles over a finite time interval \citep{haller_lagrangian_2000,shadden_definition_2005}. According to this definition, LCSs are identified as ridges (extremizing surfaces) of the FTLE field. LCSs can be further categorized into repelling and attracting LCSs, corresponding to ridges of the forward-time and backward-time FTLE that repel or attract nearby fluid particle trajectories, respectively \citep{haller_lagrangian_2011}. For a more detailed explanation of LCSs and their relation to FTLE, we refer the reader to \cite{haller_lagrangian_2015} and \cite{shadden_definition_2005}.

The largest forward-time FTLE  at position $\textbf{X}$ is defined as 
\begin{equation}
    \Lambda^{t}_{t_0}(\textbf{X}) = \frac{1}{|t - t_0|}\text{ln}\sqrt{\lambda_{\text{max}}(\textbf{C}^{t}_{t_0}(\textbf{X}))}, 
    \label{largest_ftle}
\end{equation}
while the smallest forward-time FTLE at position $\textbf{X}$ is defined as
\begin{equation}
    \Gamma^{t}_{t_0}(\textbf{X}) = \frac{1}{|t - t_0|}\text{ln}\sqrt{\lambda_{\text{min}}(\textbf{C}^{t}_{t_0}(\textbf{X}))}, 
    \label{smallest_ftle}
\end{equation}
where $\textbf{C}^{t}_{t_0}(\textbf{X}) = (\textbf{F}^{t}_{t_0}(\textbf{X}))^T\textbf{F}^{t}_{t_0}(\textbf{X})$ is the (forward) right Cauchy-Green deformation tensor, $()^T$ denotes the transpose, and $\lambda_{\text{max}}(\textbf{C}^{t}_{t_0}(\textbf{X}))$ and $\lambda_{\text{min}}(\textbf{C}^{t}_{t_0}(\textbf{X}))$ denote the largest and smallest eigenvalues of $\textbf{C}^{t}_{t_0}(\textbf{X})$,  respectively. 
Note that the eigenvalues of $\textbf{C}^{t}_{t_0}(\textbf{X})$ are always real and positive,  since $\textbf{C}^{t}_{t_0}(\textbf{X})$ is a symmetric positive definite tensor. 
Thus, the key quantity necessary for computing the FTLE fields is the deformation gradient or Jacobian of the flow map.

In Lagrangian approaches to compute the FTLE fields, a particle tracking procedure is typically adopted \cite{ONU201526}.
In this approach, particles are seeded at mesh positions, and their Lagrangian equation of motion
\begin{equation}
    \frac{\partial \textbf{x}(\textbf{X}, t)}{\partial t} \Big |_\textbf{X}
     = 
     \textbf{v}(\textbf{x}(\textbf{X}, t),t)
    \label{fluid_trajectory}
\end{equation}
is integrated in time to move particles to their final positions at time $t$ that may not coincide with a grid point. To follow the particle trajectories, a velocity interpolation algorithm is required to define the right-hand side of the above equation from the Eulerian 
velocity field. The deformation gradient is then computed typically at the initial particle position $\textbf{X}$ (that coincides with the mesh) from which the largest FTLE field or simply the FTLE field is computed using Eq.~(\ref{largest_ftle}). Note that the forward or backward FTLE fields are obtained by integrating the equation forward or backward in time, respectively.

Alternatively, the deformation gradient or Jacobian of the flow map can be obtained using only Eulerian quantities through the reference map technique discussed previously. In such approaches, the current locations of fluid particles (tracers), $\textbf{x}$, are always regarded as the coordinates of the mesh points. $\boldsymbol{\xi}(\textbf{x},t)$ then returns the take-off position $\textbf{X}$ of the tracer that has arrived at $\textbf{x}$ at time $t$. As $\textbf{X} = \boldsymbol{\xi} (\textbf{x}, t)$ is treated as continuous variables transported through the advection equation, $\boldsymbol{\xi} (\textbf{x}, t)$ does not produce exact mesh coordinates even if the tracers took off from the mesh points. Note that in this Eulerian approach, the velocity field is only required at the existing fixed Eulerian grid points (or flux points), and no additional velocity interpolation to particle locations is needed.
 

In this study, our focus is on analyzing the backward-time FTLE only, the ridges of which are identified as candidates for attracting LCSs. Attracting LCSs typically correspond to the coherent structures observed in flow visualization or identified with Eulerian diagnostics \citep{green2007, kasten2010lagrangian}. The backward-time FTLE field is computed in the same manner as the forward-time one, but now using the inverse flow map (reference map), as described below.
\newline 


\begin{itemize}
    \item Initialize the reference map variable $\boldsymbol{\xi}(\textbf{x},t=t_0) = \textbf{X}$ in the region of interest. 

    \item Advect the reference map variable using the flow field and Eq.~(\ref{eq:rmt_update}) to the final time $t$.  Now, the reference map variable describes the backward flow map $\boldsymbol{\xi}(\textbf{x},t) = \textbf{X}$.
    \item Compute the Jacobian of the backward flow map using the reference map:
    \begin{equation}
        \textbf{F}^{t_0}_{t}(\textbf{x}) = \frac{\partial \boldsymbol{\xi}(\textbf{x},t)}{\partial \textbf{x}}. 
    \end{equation}
    \item Compute the (backward)  right Cauchy-Green deformation tensor: \begin{equation}
        \textbf{C}^{t_0}_{t}(\textbf{x}) =
        ( \textbf{F}^{t_0}_{t}(\textbf{x}) )^T
        \textbf{F}^{t_0}_{t}(\textbf{x}). 
    \end{equation}
    \item Compute the largest backward-time FTLE $\Lambda^{t_0}_t (\textbf{x})$ 
    in the current configuration: 
    \begin{equation}
    \Lambda^{t_0}_{t}(\textbf{x}) = \frac{1}{|t - t_0|}\text{ln}\sqrt{\lambda_{\text{max}}(\textbf{C}^{t_0}_{t}(\textbf{x}))}.
\end{equation}
    
\end{itemize}
\ \\ \newline
\textit{Remark:}\\
Note that the backward-time FTLE is defined in the current configuration (\textbf{x}), which can be directly computed at the current position (grid points) using the reference map technique. Conversely, the forward-time FTLE is, in principle, defined in the reference configuration \textbf{X}, which can be directly computed using a Lagrangian particle approach. However, the reference map technique can be used to compute the image, under the forward flow map, of the FTLE field in the current configuration.
This can be shown by inverting \textit{Proposition 1} from \cite{haller_lagrangian_2011} to find
\begin{equation}
    \Lambda^{t}_{t_0}(\textbf{X}) = - \Gamma^{t_0}_t(\boldsymbol{\chi}^{t}_{t_0}(\textbf{X})), 
    \label{eq:ftle_duality}
\end{equation}
which shows that the largest forward-time FTLE 
can be obtained from  the smallest backward-time FTLE defined at the current position $\textbf{x}$. Proof of this equation is provided in Appendix \ref{app:duality}. 



\subsection{\label{sec:theory_tracking}Lagrangian tracking of material surface}
In the original work by \cite{kamrin_reference_2012}  which introduced the reference map technique for fluid/structure interaction, the reference map was used primarily  for computing the solid stress tensor, and the material interface was tracked separately by integrating the level-set equation. Later, \cite{valkov_eulerian_2015} observed that a level set field for the interface could be constructed directly from the reference map. \cite{jain_conservative_2019} and \cite{rycroft_reference_2020} used this to eliminate the need for separately integrating the level set equation, observing that the reference map provides a direct mapping to the initial positions of the advected particles. This feature allows for convenient updates of the level-set function or material interface directly from the reference map. \color{black}
Adopting this concept, particularly from Eq.~(25) in \cite{jain_conservative_2019}, the reference map can be demonstrated to offer basic capabilities for Lagrangian tracking of material surfaces in purely fluid flows.
For instance, in two-dimensional flows, consider a scalar field constructed from the reference map
\begin{equation*}
F(x,y, t) = \xi_1(x,y, t)^2 + \xi_2(x,y, t)^2. 
\end{equation*}
Then, the isocontour of $F(x, y, t) = R^2$ represents a material line that formed a circle at the initial time $(t=t_0)$, centered at the origin with a radius $R$. The evolution of the material points on this circle obeys the advection equation for the reference map: Each Lagrangian particle initially on the circle travels while keeping their identity fixed (or remembering where they came from), the information of which can be retrieved from the reference map. 
The isocontour of $F = R^2$ at a later time $t$ is then the collection of points that belonged to the circle at $t=0$, i.e., the Lagrangian map of the initially circular material line to its current configuration. Similarly, in three-dimensional flows, an isosurface of $\xi_1^2 + \xi_2^2$ represents the Lagrangian evolution of the material surface that formed a cylinder (tube) (whose axis is along the $x_3$ direction) at $t=0$. Other initial shapes placed at arbitrary locations are possible (e.g., plane, sphere, ellipsoid, or torus) by considering different contour functions in the form of $F(\xi_1, \xi_2, \xi_3)$ = constant. 
This property of the reference map allows for a relatively simple way to track the Lagrangian evolution of material lines or surfaces with specified initial shapes. This may be useful, for instance, for improved understanding of the Lagrangian process of vortex formation or interaction. 

Note that a similar idea of computing the passive scalar field at the current time from a direct mapping of the initial condition was explored in \cite{yang2010multi}  to conduct a multi-scale geometric analysis of Lagrangian structures in isotropic turbulence. However, the mapping was based on explicit integration of tracer particles backward in time (reminiscent of the conventional approach for computing attracting LCS) rather than carrying the Eulerian mapping to the tracers' initial positions (the reference map) forward in time. 








\section{\label{sec:numerics}Numerical details}
We turn our attention to the implementation of Eq.~(\ref{eq:rmt_update}), the passive scalar advection equation 
for the reference map $\boldsymbol{\xi}$ (or strictly, its components $\xi_{1}$, $\xi_{2}$ and $\xi_{3}$, which are scalar quantities). 
We choose to solve the conservative form of Eq.~(\ref{eq:rmt_update}) \color{black} following  \cite{kamrin_reference_2012} \color{black}, 
\begin{equation}
    \frac{\partial}{\partial t}(\rho\boldsymbol{\xi}) + \nabla\cdot(\rho\boldsymbol{\xi}\otimes\textbf{v}) = \bm{0}, 
    \label{eq:RM_conservative}
\end{equation}
which can be shown  valid for both incompressible and compressible flows 
thanks to the continuity equation. 
For the two-dimensional incompressible Taylor-Green vortex examined in Sec.~\ref{sec:result_2DTG} where the velocity field is defined analytically, the above equation (with $\rho$ omitted) is 
 discretized in space with the finite volume method,  using a second-order accurate central scheme for reconstruction of the surface flux terms. A hand coded RK3 method is used for time integration. 
  For the turbulent channel case in Sec.~\ref{sec:TCF} where
  the velocity and reference map are integrated simultaneously, a cell-centered unstructured finite-volume compressible flow solver is used.   As in \cite{jain_conservative_2019}, 
  Eq.~(\ref{eq:RM_conservative}) was discretized with the second-order central scheme and RK3 for time integration, consistent with the discretization of the continuity and momentum equations for the fluid. For detailed description of the solver and the numerical methods used therein, the reader is referred to \cite{Park2016}. 
  The flow solver was used extensively for scale-resolving simulation of low-Mach number wall-bounded turbulent flows \citep{park2014improved, park2017wall, hu2023wall, hayat2023efficient, hayat2023wall}.

Equation (\ref{eq:RM_conservative}) must be paired with proper initial and boundary conditions. A natural choice for the initial condition consistent with 
the typical choice made in the particle-based approach 
is to seed the tracer particles at the mesh points (or cell centers): 
\begin{equation}
\boldsymbol{\xi}(\textbf{x},t=t_0) = \textbf{x}. 
\label{eq:RM_IC}
\end{equation}
Note that the choice of the initial condition for $\boldsymbol{\xi}$ can have significant implication in the Eulerian simulation of solid, where the equations for the momentum and reference map are two-way coupled. The above condition corresponds to an undeformed condition, whereas as different choices could be made to assume a pre-strain/stress condition at $t=t_0$ \citep{kamrin_reference_2012, jain_conservative_2019}. This context does not apply to the one-way coupled situation in the present study, where Eq.~(\ref{eq:RM_conservative}) is regarded as the tracer equation not affecting the flow in any sense. 

\begin{figure}
    \centering
    \captionsetup{width=\linewidth}
    \includegraphics[trim=100 0 100 0,clip,width=0.9\textwidth]{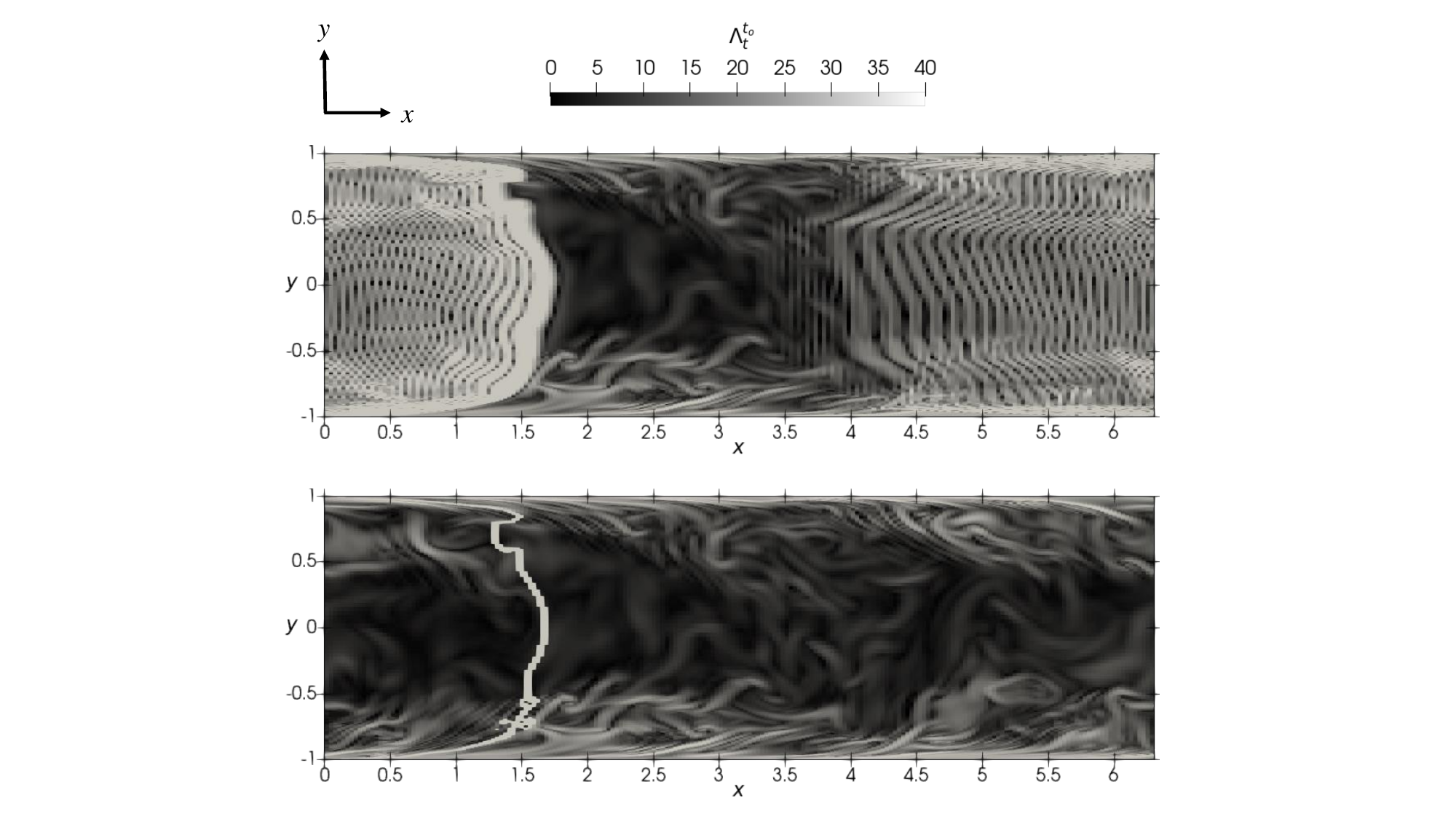}
    \caption{
    FTLE field in the $x-y$ plane at $z=1.5$ and $t^{+}=15$ obtained from the direct advection of reference map (Eq.~\ref{eq:RM_conservative}) (top),  and the advection of displacement map (Eq.~\ref{eq:disp_conservative}) (bottom).}
    \label{fig:RM_vs_displ_maps}
\end{figure}

\color{black} In this study, we consider solid walls, symmetry planes, and periodic boundary conditions. \color{black} On the boundaries where $u_n = 0$ is expected (here $u_n$ is the boundary-normal velocity) such as the solid wall or symmetry planes, the advective flux of $\boldsymbol{\xi}$ can be set to zero \color{black}($\xi_i u_n = 0$, $i=1,2,3$)\color{black}. 
The periodic boundary condition, which is straightforward to implement for the flow variables, is problematic for the reference map, especially for $\xi_1$, where $x_1$ corresponds to the main flow direction. When Eq.~(\ref{eq:RM_IC}) is used for the initial condition, the reference map variables are linear functions of $\textbf{x}$, and periodicity cannot be enforced inherently. 
\color{black}
Advection of such discontinuous data with central schemes can be problematic, because 
the dispersion error may lead to the formation of wiggles and their spread along or opposite to the flow direction, as shown in figure~\ref{fig:RM_vs_displ_maps} for $Re_\tau = 180$ turbulent channel. 
This is often accompanied with violation of the boundedness 
for passively advected scalars, where the reference map at all times must be bounded by its range from  the initial condition. 
This can be remedied by rewriting  Eq.~(\ref{eq:RM_conservative}) 
in terms of the displacement map 
\begin{equation}
    \boldsymbol{g}(\textbf{x}, t) \equiv
    \textbf{x} - \textbf{X}(\textbf{x}, t) = 
    \textbf{x} - \boldsymbol{\xi}(\textbf{x}, t), 
    \label{eq:displacement}
\end{equation}
which is the difference between 
the current position of particles arriving at the grid points (\textbf{x}) 
and its initial position (\textbf{X}). 
The 
 equations for the displacement map in the conservative or material forms 
 are 
\begin{equation}
    \frac{\partial}{\partial t}(\rho\boldsymbol{g}) + \nabla\cdot(\rho\boldsymbol{g}\otimes\textbf{v}) = \rho \textbf{v}, 
    \ \ \ \text{or} \ \ \ 
        \frac{D\boldsymbol{g}}{D t} = \frac{\partial\boldsymbol{g}}{\partial t} + \textbf{v}\cdot\nabla\boldsymbol{g} = \textbf{v}. 
    \label{eq:disp_conservative}
\end{equation}
Note that, unlike the reference map,  periodicity of the displacement map is guaranteed, provided that its initial condition and the velocity field are periodic. 
Fortunately, from Eq.~(\ref{eq:RM_IC}), the initial condition for the displacement map is a zero field  which is periodic. 
 Advection of the displacement field therefore does not suffer from the dispersion error. The strategy for periodic boundary conditions then will be 
to update the displacement field via Eq.~(\ref{eq:disp_conservative}), and to recover the reference map via Eq.~(\ref{eq:displacement}). Note that $\boldsymbol{g}$, especially its  streamwise component, will in general grow/decay in time due to the right-hand side of Eq.~(\ref{eq:disp_conservative}), and the reference map obtained from Eq.~(\ref{eq:displacement}) may 
indicate initial particle locations outside of the computational domain. In the spirit of periodicity, this can be corrected by keeping adding/subtracting the spatial period to/from the reference map until it falls within the domain. 
While the reference map so obtained is free from oscillations, it does exhibit a discontinuity front that is advected in the flow, potentially impacting the computation of derived quantities, such as the FTLE field as shown in the bottom plot in figure \ref{fig:RM_vs_displ_maps}.
Note that it is still easy to 
identify extensive regions of the flow that \color{black} do \color{black} not contain the discontinuity. 

Although not explored in the present  study, on the outflow boundary zone where the flow is expected to exit normal to the boundary, 
one may choose to treat the boundary data as unknown and implement a one-dimensional convective outflow boundary condition in the boundary normal direction \citep{schluter2005outflow}, which highly resembles the equation for $\boldsymbol{\xi}$. 

\color{black}

\begin{figure}
    \centering
    \captionsetup{width=\linewidth}
    \includegraphics[width=0.6\textwidth]{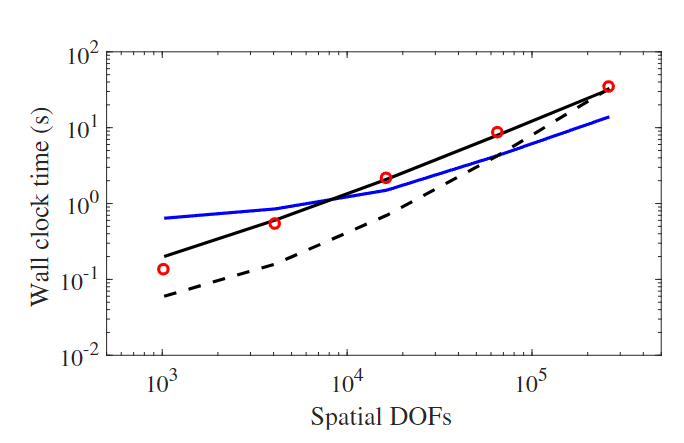}
    \caption{
    Comparison of the FTLE computation cost on MATLAB for the 2D Taylor-Green vortex (Sec.~\ref{sec:result_2DTG}). 
    Blue solid line, particle-based approach (LCS Tool); 
    red circles, asymptote for linear scaling; 
    black solid line, reference map approach where 
    the time step was fixed to $dt$=2e-2, corresponding to the maximum CFL condition on the finest grid; 
    black dashed line, reference map approach with the fixed CFL number (1.6). 
    The effect of the time step on the computed FTLE field was seen to be \color{black} negligible \color{black} in this test case. LCS Tool is a vectorized code, whereas the reference map code is not. 
    }
    \label{fig:cost}
\end{figure}

In concluding remarks on computational cost, it is noted that the integration of the reference map equations on the fly, along with the Navier-Stokes equations, incurs a 40\% increase in wall clock time compared to standalone flow computations. However, we note  that the integration of reference map variables is necessary only for a limited time interval, usually after the velocity field reaches a statistically steady state.
For the computation of the FTLE field and material surface tracking in channel flow at $Re_\tau = 180$ (Sec.~\ref{sec:TCF}), we find that a time duration of 20 in viscous wall units and a small fraction of the large-eddy turnover time ($\delta/u_\tau$) are sufficient, respectively, to obtain the converged results (30 to 40 viscous time units were  suggested for the same flow by \cite{green2007} and \cite{rockwood2018tracking}). 
Nevertheless, it is acknowledged that the additional overhead is considerably higher compared to Eulerian diagnostics, such as the $Q$ or $\lambda_2$ criteria, for coherent structure detection. 
When the velocity field is pre-computed or accessible analytically, the cost of the present approach for FTLE computation only (without solving the Navier-Stokes) can be compared to that of the standard particle approach. Figure \ref{fig:cost} presents this comparison for the test case in Sec.~\ref{sec:result_2DTG}. With a fixed time step, the cost of the reference map technique scales linearly with the number of grid points, as expected. The standard particle approach (LCS tool, \cite{ONU201526}), employing a vectorized explicit high-order adaptive ODE solver with variable time steps, likely uses the maximum time steps allowed for stable time integration. Under a similar constraint, the reference map approach appears to be considerably faster than the particle approach up to $10^5$ grid points. Beyond this point, the particle approach is observed to be more efficient, presumably due to the substantial speedup brought by optimal memory access patterns in vectorized codes.

\section{\label{sec:result}Results and discussions}

\subsection{\label{sec:result_2DTG}2D TG vortex}

We consider incompressible 
two-dimensional steady (inviscid) Taylor-Green vortex 
to examine the accuracy of FTLE calculation based on reference map. We will also elucidate the idea of using  reference map for material line tracking. The flow is defined in a doubly periodic domain between 0 and 2$\pi$, where  the velocity field is given by $u(x,y) = \sin x \cos y$ 
and $v(x,y) = -\cos x \sin y$. \color{black} Note for this example, we consider non-dimensional units\color{black}.
At $t=0$, the reference map variables are initialized with the coordinates of grid collocation points, corresponding to the initial seeding of tracer particles on the grid points. Note that the velocity component normal to the computational boundaries is 0, and the boundaries are effectively non-penetrating walls.

\begin{figure}
    \centering
    \captionsetup{width=\linewidth}
    \includegraphics[width=1\textwidth]{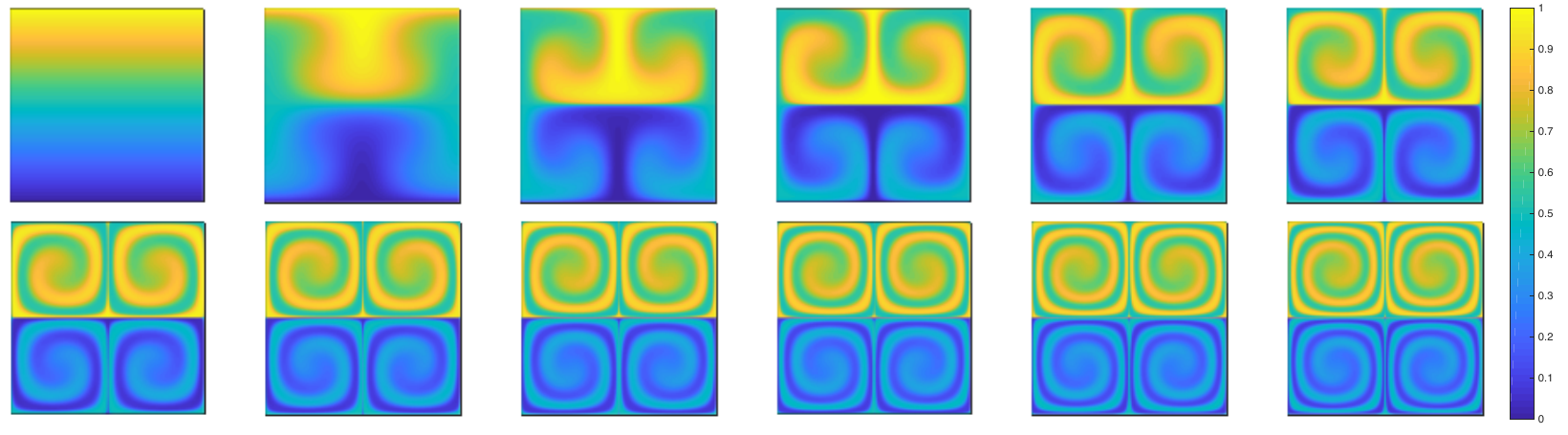}
    \caption{2D steady Taylor-Green Vortex: 
    Time evolution of contour of $x_2$-component of the reference map 
    ($\xi_2(x,y,t)/(2\pi)$) \color{black} on the periodic domain $[0,2\pi]^2$  from $t=0$ to $t = 10$ in the top row and $t=12$ to $t=22$ in the bottom row \color{black} (time interval =2). 
    }
    \label{fig:2D_TG_x2}
\end{figure}

\begin{figure}
    \centering
    \captionsetup{width=\linewidth}
    \includegraphics[width=0.9\textwidth]{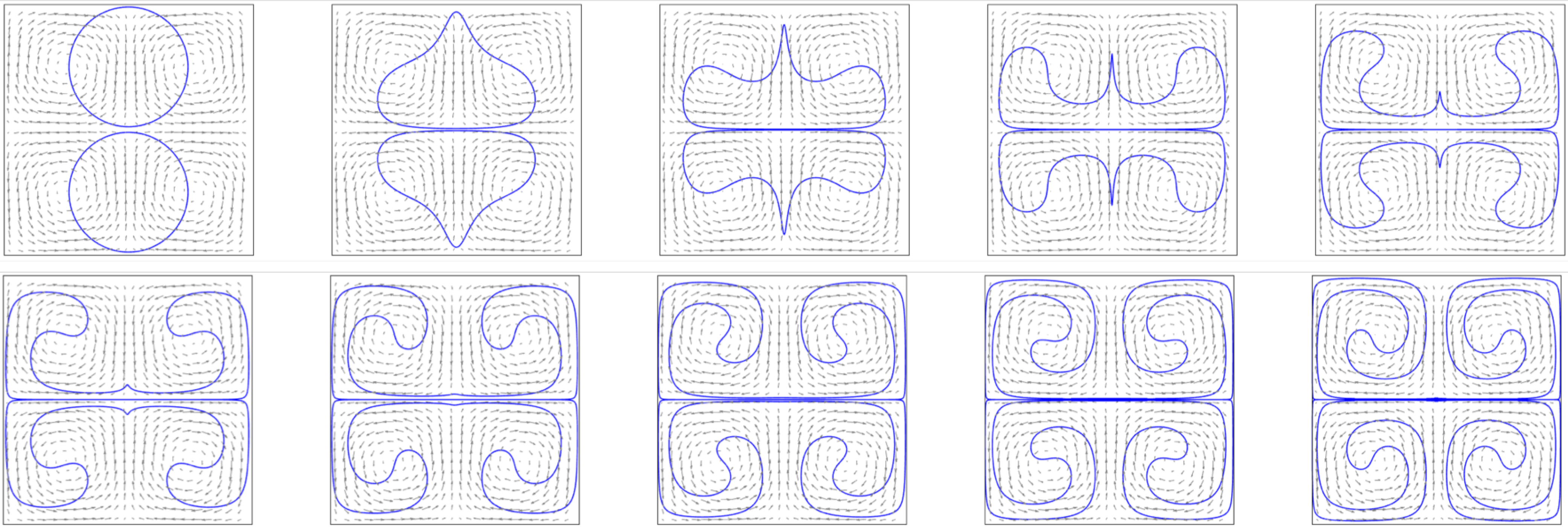}
    \caption{2D steady Taylor-Green Vortex: 
    Time evolution of two initially circular material lines
    from $t=0$ (top left) to $t=9$ (bottom right) (time interval = 1). 
    Material lines are tracked from iso-level lines of two scalar functions defined from the reference map: 
    $F_1(x,y,t) = (\xi_1(x,y,t) - \pi)^2 + (\xi_2(x,y,t) - 1.5\pi)^2$ and $F_2(x,y,t) = (\xi_1(x,y,t) - \pi)^2 + (\xi_2(x,y,t) - 0.5\pi)^2$. 
    The circles are tracked with $F_1 = 1.5^2$ and  $F_2 = 1.5^2$. 
    }
    \label{fig:2D_TG_material_line}
\end{figure}

Figure \ref{fig:2D_TG_x2} shows the time evolution of the color contour of 
$\xi_2(x,y,t)$ from $t=0$ to $t=22$ \color{black} computed on a $512\times512$ uniformly spaced grid \color{black}. 
Note that the initial condition for $\xi_2$ is the $y$ coordinates of the tracers placed on the grid points at $t=0$. Therefore, the line (or region, roughly speaking) with identical color  can be identified as 
a timeline (or material/Lagrangian area). The 
roll up of the initially horizontal material lines/surfaces into spiral ones  
due to the background vortical flows is observed. 
While the background flow is steady, the contour of $\xi_2$ does not reach 
the steady state,  because the tracers' current position  keeps varying in time. 
Figure \ref{fig:2D_TG_material_line} shows evolution of two material lines, which form circles at $t=0$ separated by a distance 0.6$\pi$ along the vertical centerline at $x = \pi$. 
 Consistent with the background vortices and the evolution of $\xi_2$ from figure~\ref{fig:2D_TG_x2}, each circular material line first  migrates to the center of the domain and deforms into a \color{black}Hershey's \color{black} Kisses shape,  squeezes into a 
  pancake shape by the background shear, eventually rotating in a spiral pattern in each quadrant.

\begin{figure}
    \centering
    \captionsetup{width=\linewidth}
    \includegraphics[width=0.9\textwidth]{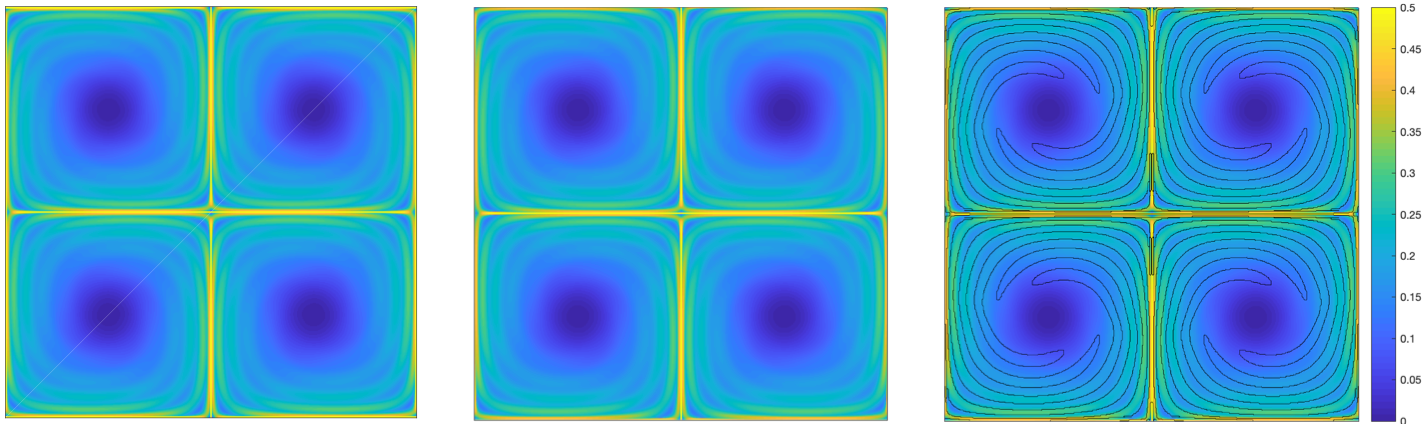}
    \caption{2D steady Taylor-Green Vortex: Comparison of color contours of the largest backward-time FTLE ($\Lambda^{t_0}_t$) obtained from $t=10$ to $t_0=0$. Left, from LCS Tool (particle-based approach). Middle, from the reference map (present study).
    Right, Same as Middle, but the solid lines are the FTLE ridges as identified from Eq.~(\ref{eq:attracting_LCS}). 
    Grid/particle resolutions are the same in all cases (\color{black} $512\times 512$\color{black}). 
    }
    \label{fig:2D_TG_ftle}
\end{figure}

Figure \ref{fig:2D_TG_ftle} compares the backward-time FTLE 
($\Lambda^{t_0}_{t}(\textbf{x})$) obtained with the reference map to that obtained with a direct particle-based approach.
\color{black}
For LCS Tool, the main particle grid for Lagrangian particle tracking is identical to the grid for the reference map (a $512\times512$ uniformly-spaced grid). 
However, LCS Tool utilizes an auxiliary particle grid (4 points surrounding each main grid point) 
 with the  spacing of 1–10\% of the main grid spacing to 
 approximate the deformation gradient accurately with finite difference  \citep{ONU201526}. 
\color{black}
It is readily confirmed from the left and middle panels of figure~\ref{fig:2D_TG_ftle} that the FTLE fields produced by the particle-based method  and the reference map approach are indistinguishable, which serves as validation of the current approach. 
As ridges (local maximizing lines/surfaces) of the backward-time FTLE are candidates for attracting material lines or attracting LCS \citep{haller_lagrangian_2000}, we further attempt to locate the FTLE ridges through the zero-gradient  contours of the FTLE 
that satisfy a second-derivative condition in the $x$ direction 
\begin{equation}
\text{Attracting LCS: \ \ }
| \nabla \Lambda^{t_0}_t | = 0 
\text{ \ and \ }
\frac{\partial^2 }{\partial x^2} \Big ( \Lambda^{t_0}_t (\textbf{x}) \Big )< 0. 
\label{eq:attracting_LCS}
\end{equation}
For the visualization purpose only, this can be done simply by assigning a large negative value to the computed FTLE gradient magnitude field when 
$\frac{\partial^2 }{\partial x^2} \Big ( \Lambda^{t_0}_t \Big )$ is found locally positive  (masking out potential local minima/troughs). 
We find 
Eq.~(\ref{eq:attracting_LCS}) produces reasonable representation of the FTLE ridges (figure~\ref{fig:2D_TG_ftle}, right panel), which were identical to those obtained with the full Hessian test. 
\newline

\begin{figure}
    \centering
    \captionsetup{width=\linewidth}
    \includegraphics[width=1\textwidth]{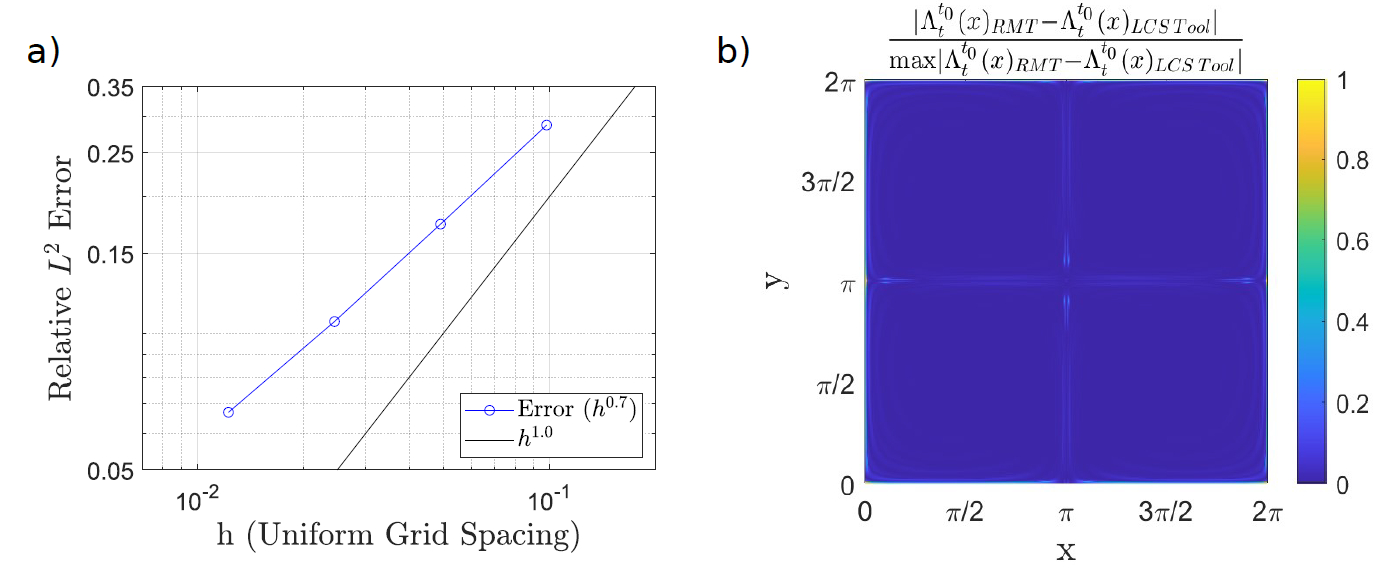}
    \caption{\color{black} 2D steady Taylor-Green Vortex: (a) Relative $L^2$ error between a between a fine grid ($1024 \times 1024$) reference backward-time FTLE field computed using the RMT approach and those on a sequence of successively refined grids, where we obtain a convergence rate of approximately $h^{0.7}$. (b) Normalized absolute difference  between the backward-time FTLE fields computed using the particle-based (LCS Tool) and the reference-map approaches on a 512$\times$512 grid.\color{black} 
    }
    \label{fig:bftle_error}
\end{figure}

\color{black} 
Additionally, we perform a convergence study of the reference map based approach under grid refinement, and take the backward-time FTLE field computed on a fine grid $(1024 \times 1024)$ using the reference map approach as a reference solution. Specifically, we consider the following uniform grids $64\times64$, $128\times128$, $256\times256$, $512\times512$, and compute relative $L^2$ errors 
against the fine-grid result. 
Figure \ref{fig:bftle_error}(a) plots these errors as a function of uniform grid spacing, demonstrating the reference map backward-time FTLE field converges at a rate of approximately 0.7. Next, we assess the difference  between the backward-time FTLE fields computed on the same grid using the reference map approach and LCS Tool. The difference, albeit small, is concentrated most in the boundaries between the neighboring spiral vortices where 
 rapid variations in the strain rate and FTLE are observed, as seen in the normalized absolute difference plot in Figure \ref{fig:bftle_error}(b). 
 As noted earlier, the use of the auxiliary particle grid results in extra accuracy in LCS Tool, which is absent in the current reference map approach.  
Nevertheless, we observed the difference decreases under grid refinement. \color{black}

\subsection{\label{sec:LDC}2D Lid-driven cavity}

\color{black} The next case we consider for the validation of reference-map-based FTLE calculation is an incompressible two-dimensional steady lid-driven cavity at a Reynolds number of 100. The geometry is a square cavity with unit edge length, consisting of three rigid walls with no-slip conditions and the top wall moving along positive $x$ with a tangential unit velocity, inducing the flow inside the cavity. At $t=0$ 
where the flow has reached the steady state, the reference map variables are initialized with the coordinates of grid collocation points. Similar to the 2D TG vortex, the velocity component normal to all the walls is 0, simplifying the boundary closure of the reference map as discussed in Sec.~\ref{sec:numerics}. 

\begin{figure}
    \centering
    \captionsetup{width=\linewidth}
    \includegraphics[trim=0 200 0 200,clip,width=0.9\textwidth]{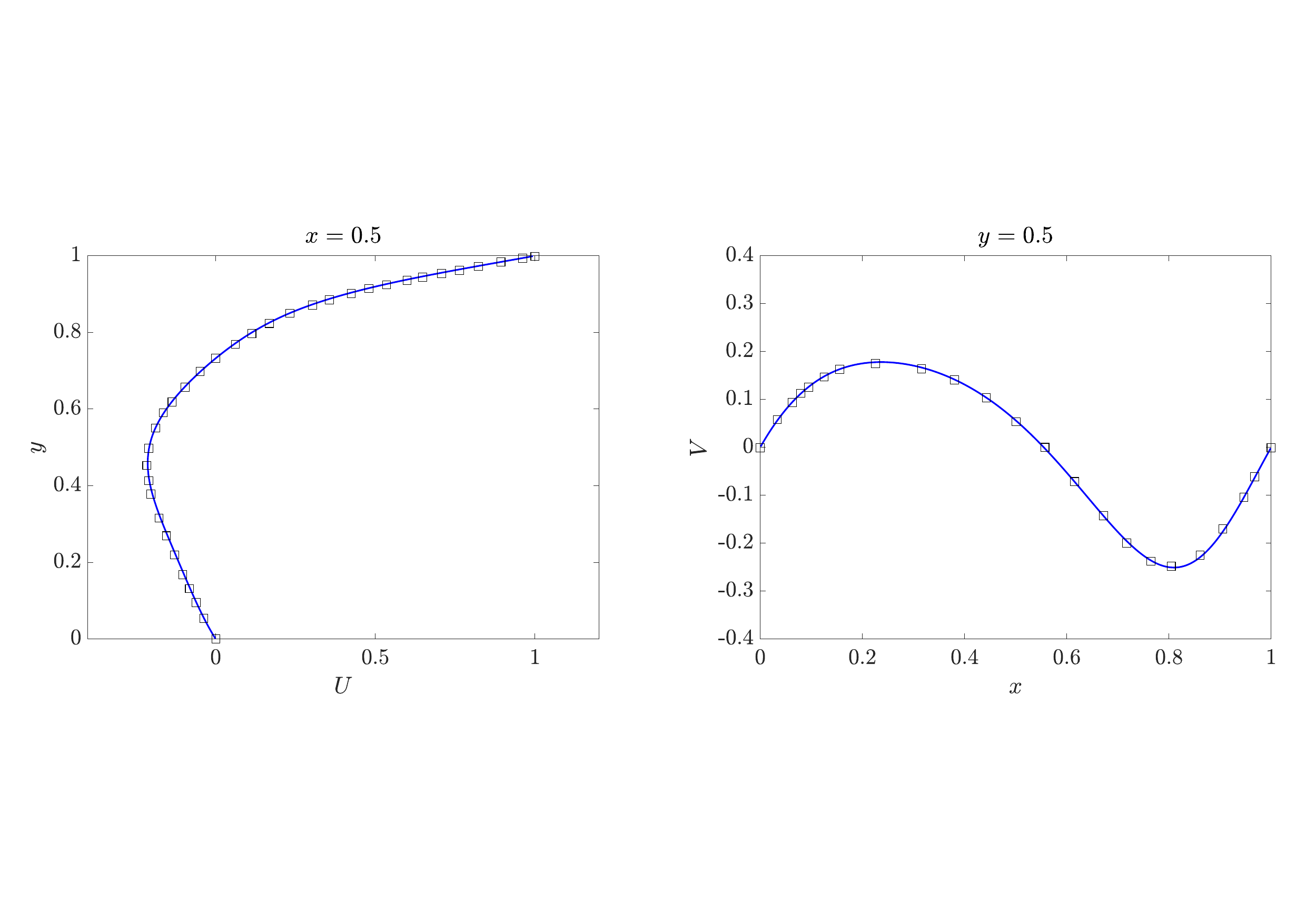}
    \caption{\color{black} Validation of velocity profiles in 2D lid-driven cavity at Re=100: horizontal velocity at $x=0.5$ (left) and vertical velocity at $y=0.5$ (right). Squares, reference data from \cite{ghia1982}; solid lines, present simulation on a 512$\times$512 grid. \color{black}
    }\label{fig:LDC_validation_velocity}
\end{figure}

\begin{figure}
    \captionsetup{width=\linewidth}
    \begin{minipage}{0.48\textwidth}
    \includegraphics[trim=120 220 140 230,clip,width=\linewidth]{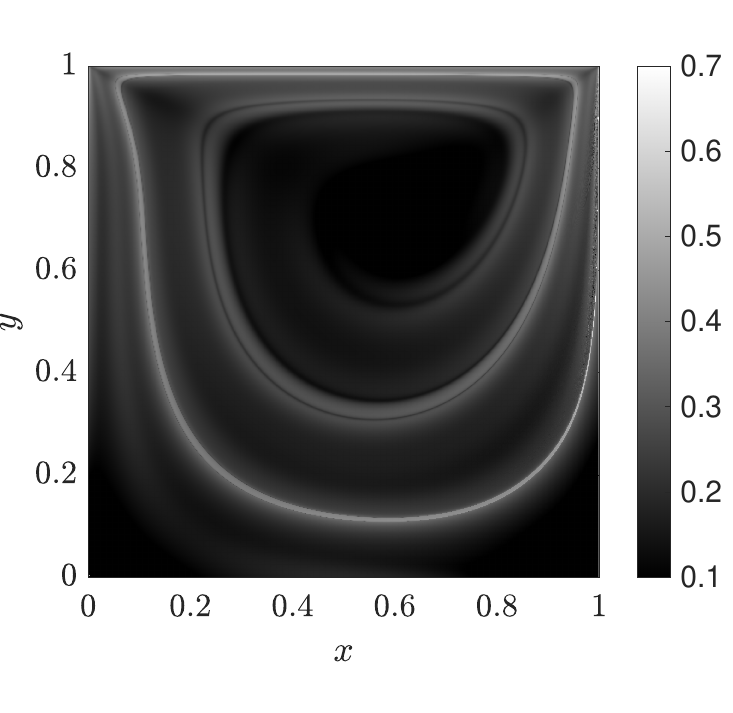}
    \end{minipage}
    \hspace{\fill} 
    \begin{minipage}{0.48\textwidth}
    \includegraphics[trim=120 220 140 230,clip,width=\linewidth]{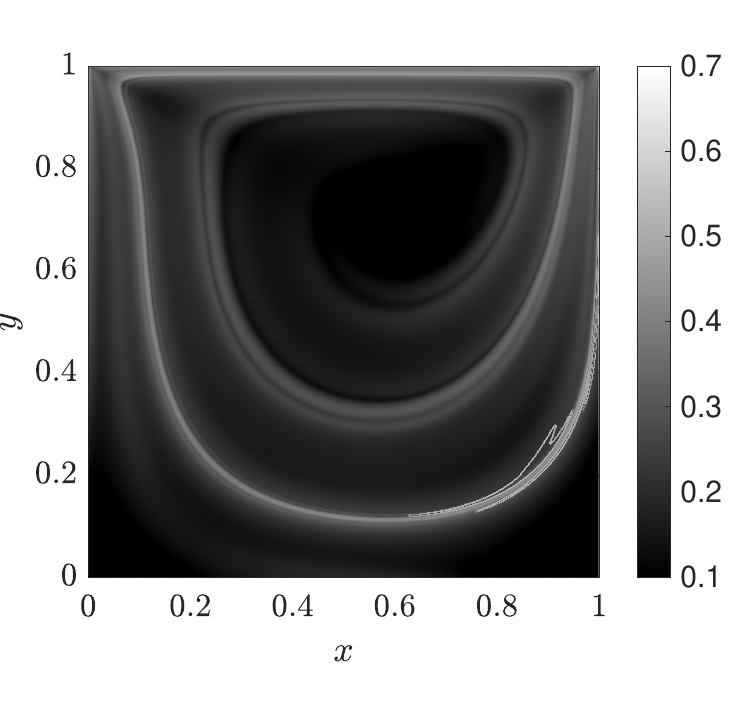}
    \end{minipage}
\caption{\color{black} 2D lid-driven cavity at Re=100: Comparison of color contours of the largest backward-time FTLE ($\Lambda^{t_0}_t$) obtained from $t=11.24$ to $t_0=0$. Left, from LCS Tool (particle-based approach). Right, from the reference map (present study). 
}\label{fig:ldc_ftle}
\end{figure}

\color{black} In Fig.~\ref{fig:LDC_validation_velocity}, the steady velocity profiles at the mid-sectional lines $x=0.5$ and $y=0.5$ from a simulation using 512$\times$512 grid are validated against the profiles from \cite{ghia1982}. The profiles of both horizontal and vertical velocity show good agreement with the reference data. Figure~\ref{fig:ldc_ftle} shows the comparison of the backward-time FTLE 
($\Lambda^{t_0}_{t}(\textbf{x})$) from the present reference-map approach and that from a direct particle-based approach using the LCS Tool described in the previous section, both computed on a $512\times512$ uniformly-spaced grid. For the Lagrangian tracking in the LCS Tool, a fine auxiliary grid is used with a grid spacing equal to 1\% of the main grid spacing. A visual inspection of the FTLE fields produced by the LCS Tool and the reference map approach in figure~\ref{fig:ldc_ftle}, shows a reasonable overall agreement between the two fields, except for some localized regions of high $\Lambda^{t_0}_{t}(\textbf{x})$ toward the bottom right of the right panel. These localized regions appear to be originating from the singularity of the boundary condition at the top-right corner \citep{bouffanais2007large, sousa2016lid, kuhlmann2019lid}. 
This effect is not observed in the velocity field but only in the FTLE field. Note that the scalar advection equation (which the reference map obeys) is more prone to the dispersion error due to such discontinuous data. 
However, we find that the local nature of these artifacts prevent them from significantly polluting most of the FTLE field, even after a long time integration of 11 time unit. \color{black} 


\subsection{\label{sec:TCF}Turbulent Channel flow}





The DNS of turbulent channel flow at $Re_{\tau} = 180$ was performed, where $Re_{\tau}$ is the friction  Reynolds number based on the channel half-height ($\delta$), friction velocity ($u_{\tau}$), and kinematic viscosity ($\nu$). A minimal channel with dimensions $x/\delta \in [0,2\pi]$ in the streamwise direction, $y/\delta \in [-1,1]$ in the wall-normal direction, and $z/\delta \in [0,\pi]$ in the spanwise direction was used, where $x$, $y$, and $z$ represent the streamwise, wall-normal, and spanwise directions, respectively. Periodicity was imposed in the streamwise and \color{black} spanwise \color{black} directions. A uniform grid was used in the streamwise and spanwise directions, whereas a stretched grid was employed in the wall-normal direction, with the number of grid points in each direction given by $(N_{x},N_{y},N_{z}) = (200,260,160)$. \color{black} The wall-normal spacing is given by the geometric progression \color{black} $\Delta y_{i} = r^{i - 1} \Delta y_{1}$, $i = 1,2,...,N_{y}/2$, where $r=1.035$ and $\Delta y_{1}/\delta = 4.04 \times 10^{-4}$. The resulting grid \color{black} spacings \color{black} in wall units are $(\Delta x^{+},\Delta y_{min}^{+},\Delta y_{max}^{+},\Delta z^{+}) \approx (6,0.07,6,4)$. \color{black} Note that the superscript ``$+$'' denotes quantities non-dimensionalized by viscous wall units, where the non-dimensionalized distance $l^+$ and time $t^+$ are given by, $l^+ = l u_{\tau} / \nu$ and $t^{+} = t u^{2}_{\tau}/\nu$, respectively\color{black}. 
After the simulation reached statistical stationarity \color{black}(which is verified through the convergence of integrated wall forces and bulk velocity)\color{black}, statistics were collected over a nondimensional time \color{black} period \color{black} $T^{+} = t u^{2}_{\tau}/\nu  \approx 9500$ ($t U_{b}/\delta  \approx 822$ \color{black}, where $U_b$ is the bulk velocity \color{black}). Profiles of the mean velocity and the Reynolds stresses from the present DNS were validated against the channel flow DNS of \citet{Kim1987}. 
\color{black} After the flow was fully developed, computation of the FTLE field was started at an arbitrary time $t=t_0$, \color{black} where the reference map variable was set to be the coordinates of cell centers. 

\begin{figure}
    \centering
    \captionsetup{width=\linewidth}
    \begin{minipage}{0.47\textwidth}
        \includegraphics[trim=0 0 80 0,clip,width=\linewidth]{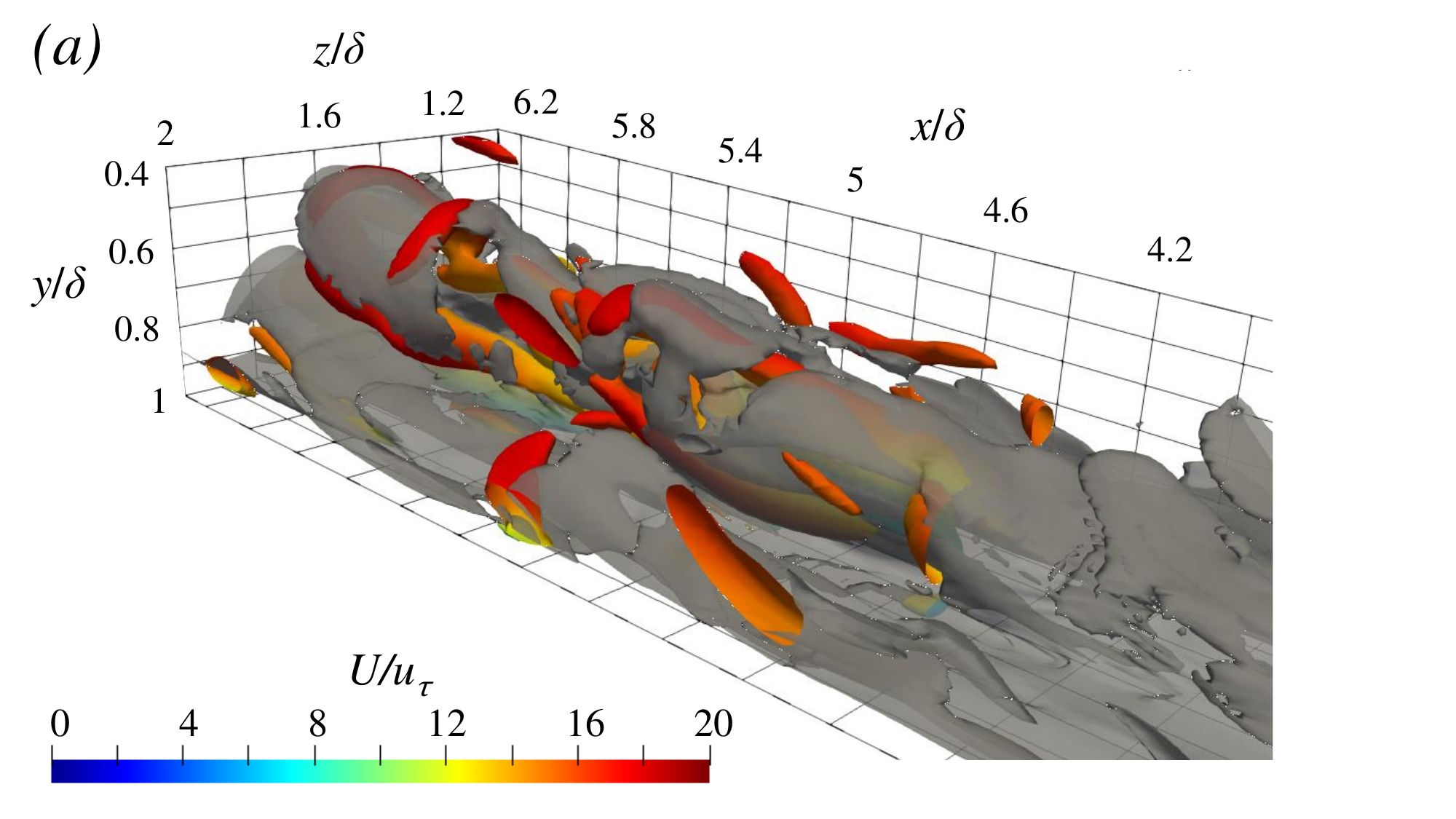}
    \end{minipage}
    \hspace{\fill} 
    \vspace*{0.5cm} 
    \begin{minipage}{0.47\textwidth}
        \includegraphics[trim=10 0 25  0,clip,width=\linewidth]{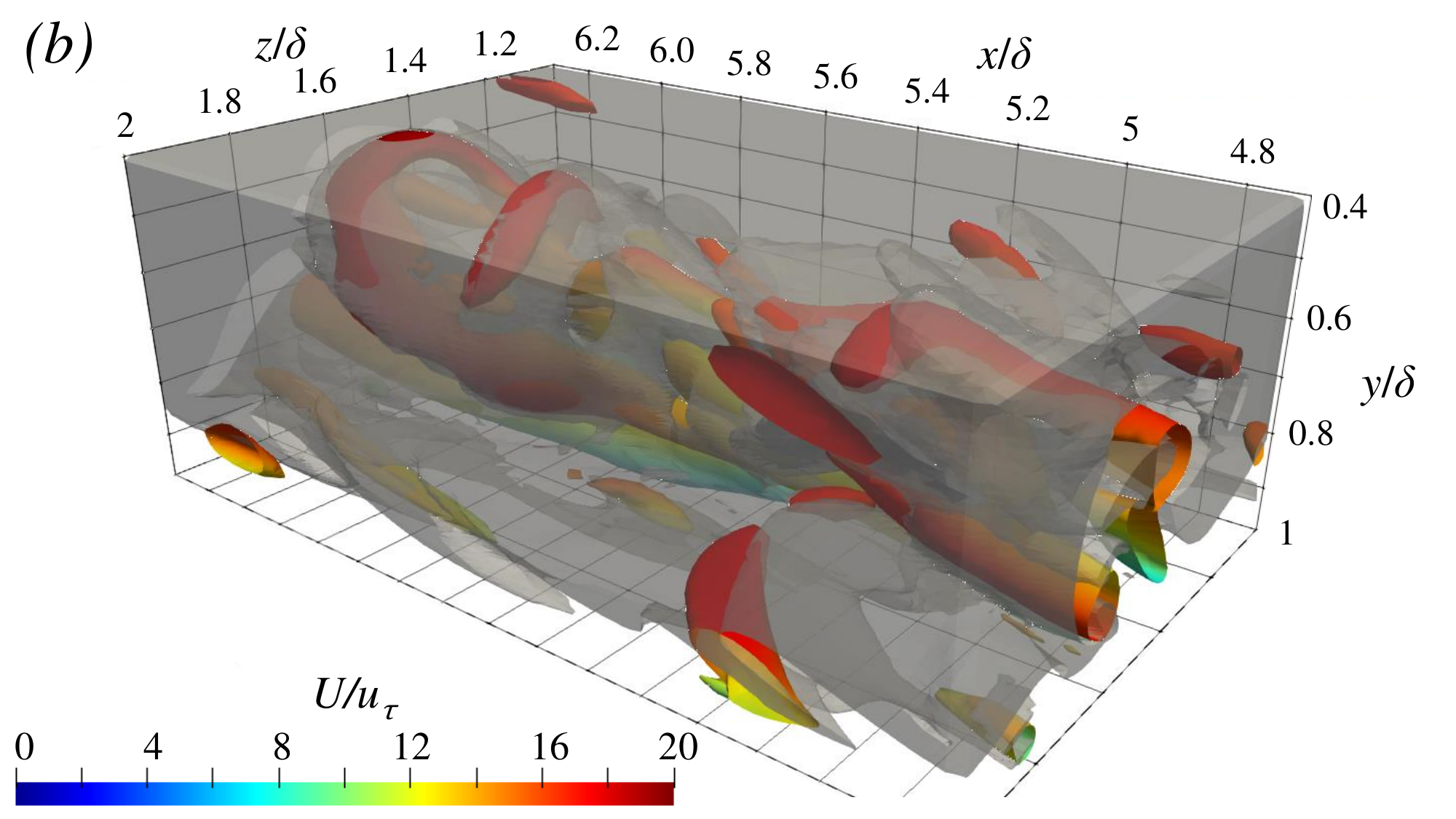}
    \end{minipage}
    \vspace*{0.5cm} 
    \begin{minipage}{0.47\textwidth}
        \includegraphics[width=\linewidth]{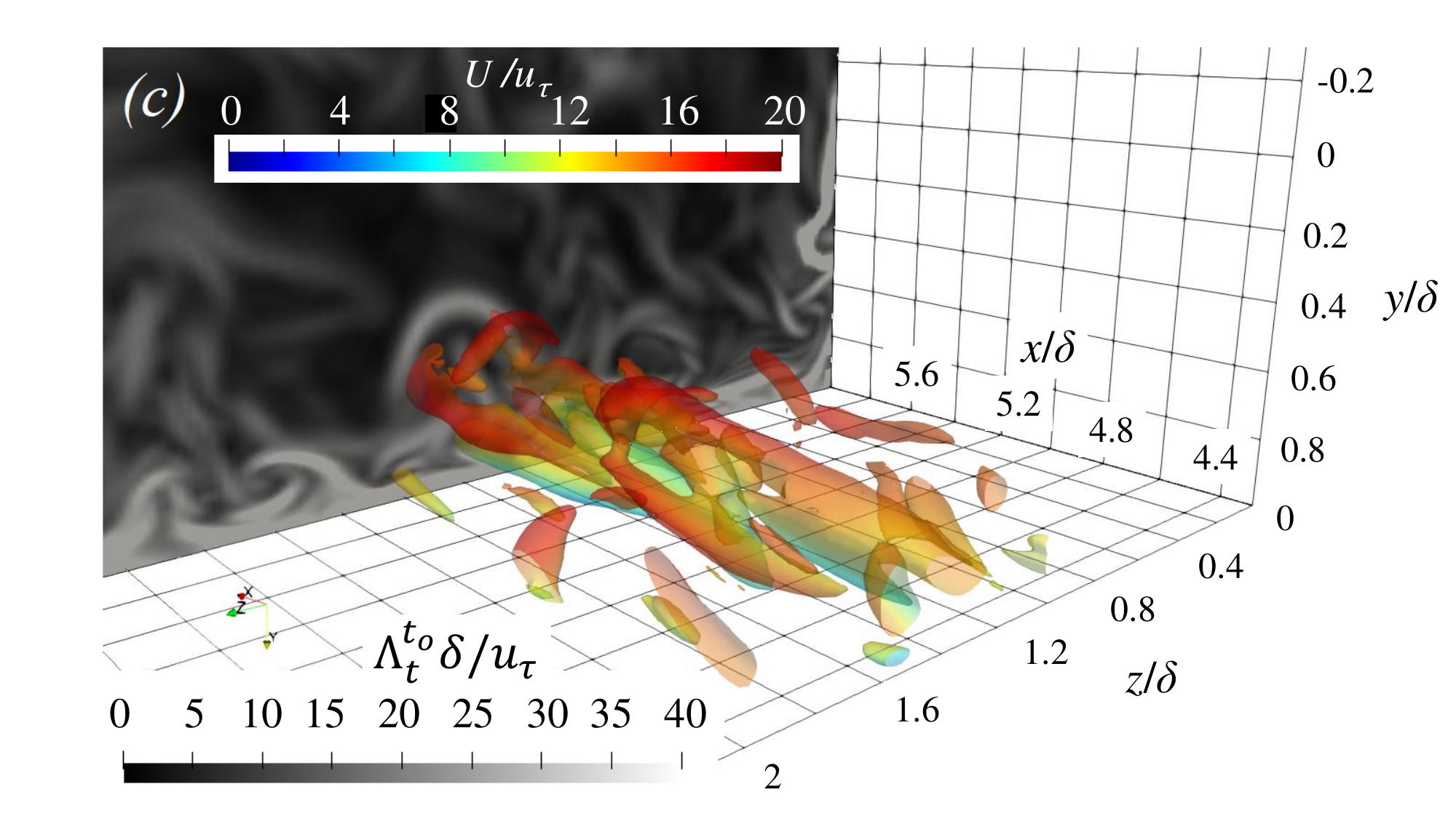}
    \end{minipage}
    \hspace{\fill} 
    \begin{minipage}{0.47\textwidth}
        \includegraphics[width=\linewidth]{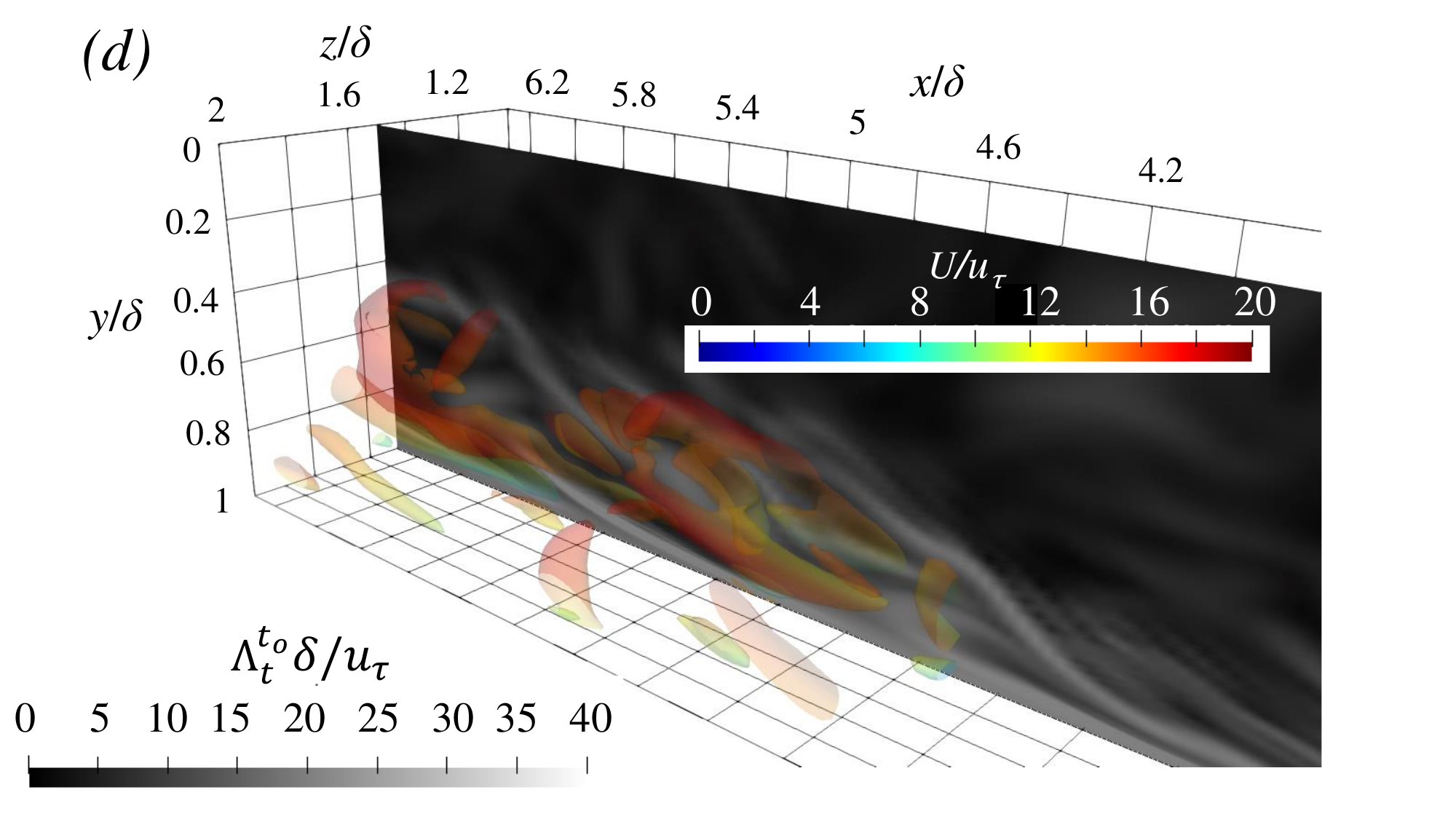}
    \end{minipage}
    \vspace*{0.5cm} 
    \begin{minipage}{0.47\textwidth}
        \includegraphics[trim=50 0 160 0,clip,width=\linewidth]{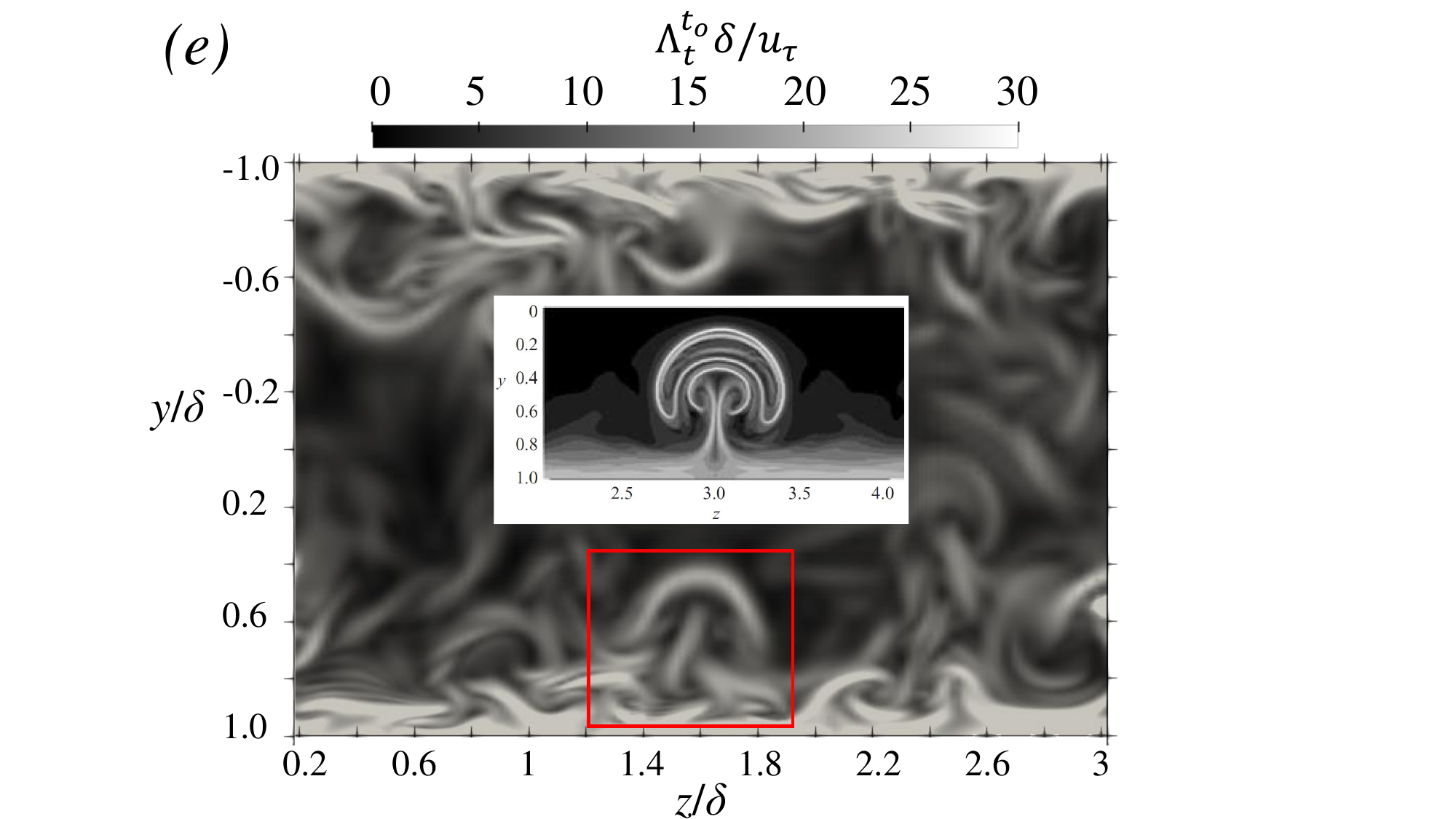}
    \end{minipage}
    \hspace{\fill} 
    \begin{minipage}{0.49\textwidth}
        \includegraphics[trim=20 55 40 10,clip,width=\linewidth]{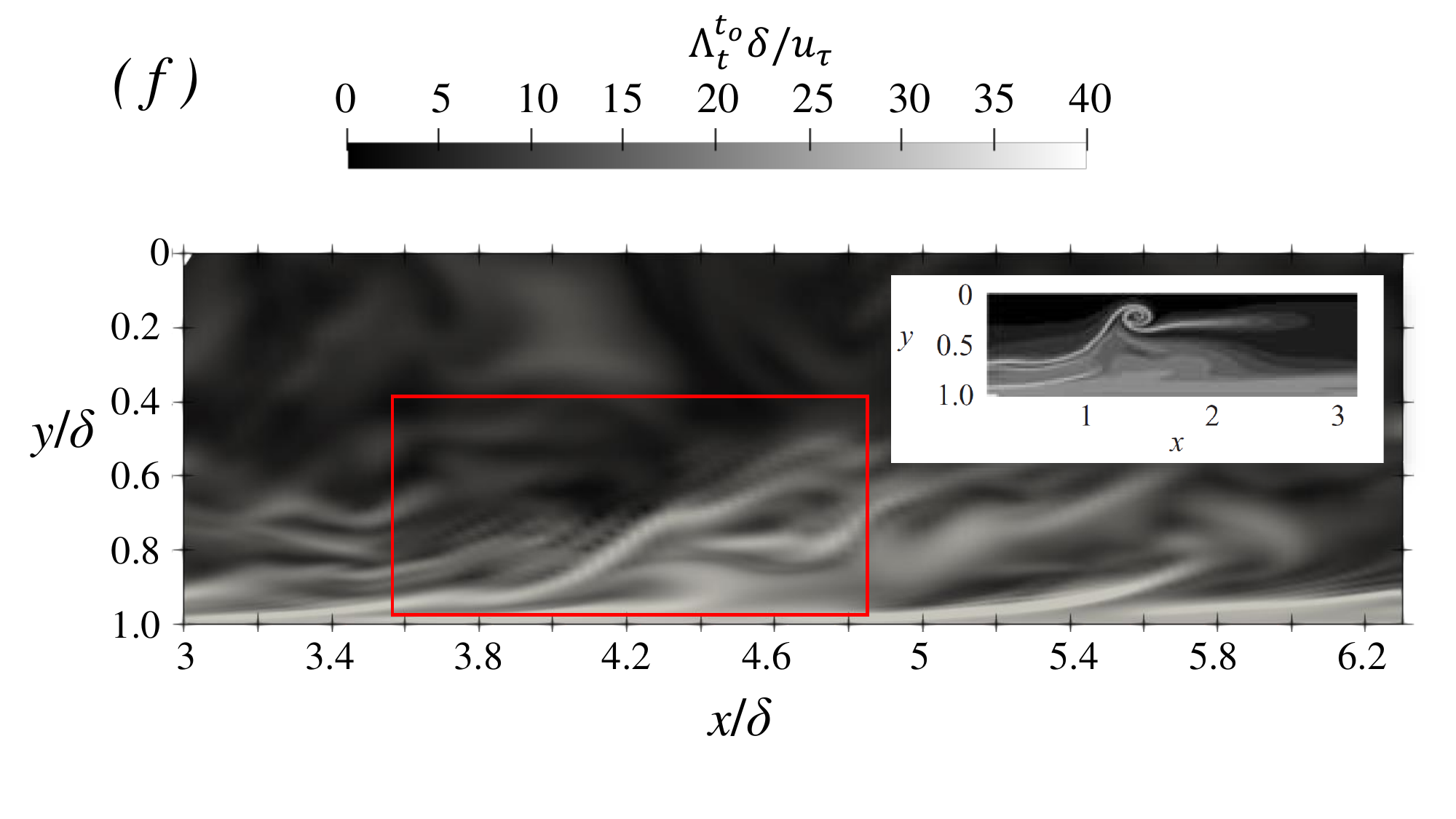}
    \end{minipage}

    \begin{minipage}{0.55\textwidth}
        \includegraphics[trim=10 0 0 0,clip,width=\linewidth]{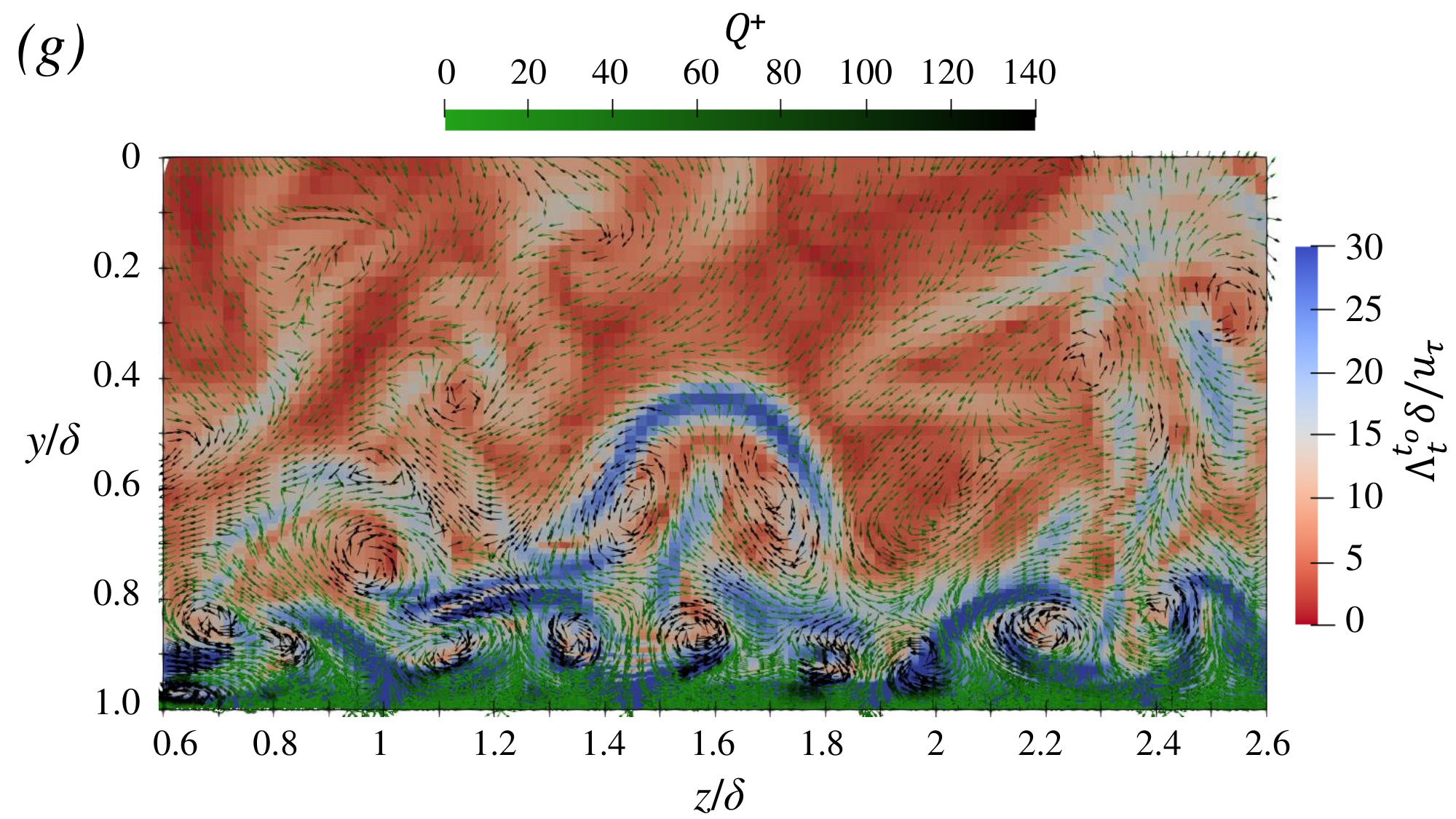}
    \end{minipage}
\caption{Comparison of the backward-time FTLE with \color{black} the $Q$-criterion \color{black} structures at $t^{+} = 12 + t_0^{+}$. In the sub-figures where available, the iso-surfaces colored by instantaneous streamwise velocity are those of $Q^+=350$, which are shown along with $(a)$ iso-surfaces of \color{black}$\Lambda^{t_0}_{t} \delta/u_{\tau}=20$\color{black}, $(b)$ iso-surfaces of \color{black}$\Lambda^{t_0}_{t} \delta/u_{\tau} \leq 15$\color{black}, $(c)$ backward-time FTLE in a constant-streamwise cut at $x/\delta=6$, $(d)$ backward-time FTLE in a constant-spanwise cut at $z/\delta=1.5$; $(e)$ \& $(g)$ show $\Lambda^{t_0}_{t}$ in the same constant-streamwise cut as $(c)$; $(f)$ shows $\Lambda^{t_0}_{t}$ in the same constant-spanwise cut as $(d)$. In $(e)$ \& $(f)$, the red boxes identify structures to be compared with those from \cite{green2007} shown in the insets. \color{black}Note that \cite{green2007} used a particle-tracking (auxiliary) grid for the FTLE calculation that was six-times finer in all directions than the flow grid for DNS\color{black}. In $(g)$, the colored arrows indicate the direction of in-plane instantaneous velocity vector (colored by \color{black} the \color{black} $Q$-criterion).}\label{fig:TCF_LCS}
\end{figure}

\begin{figure}
    \captionsetup{width=\linewidth}
    \begin{minipage}{0.58\textwidth}
    \includegraphics[trim=0 0 0 0,clip,width=\linewidth]{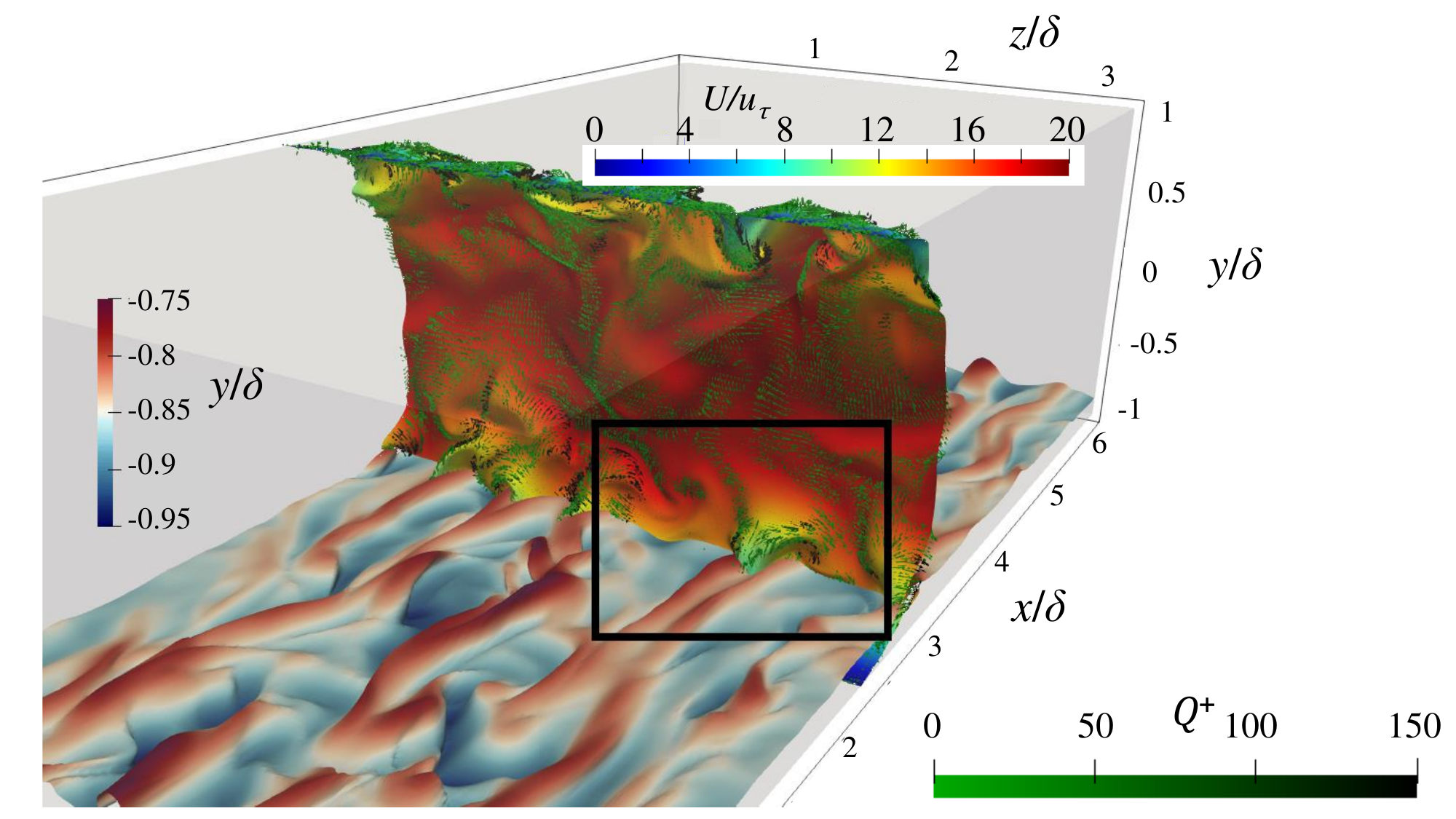}
    \end{minipage}
    \hspace{\fill} 
    \begin{minipage}{0.4\textwidth}
    \includegraphics[trim=250 50 120 0,clip,width=\linewidth]{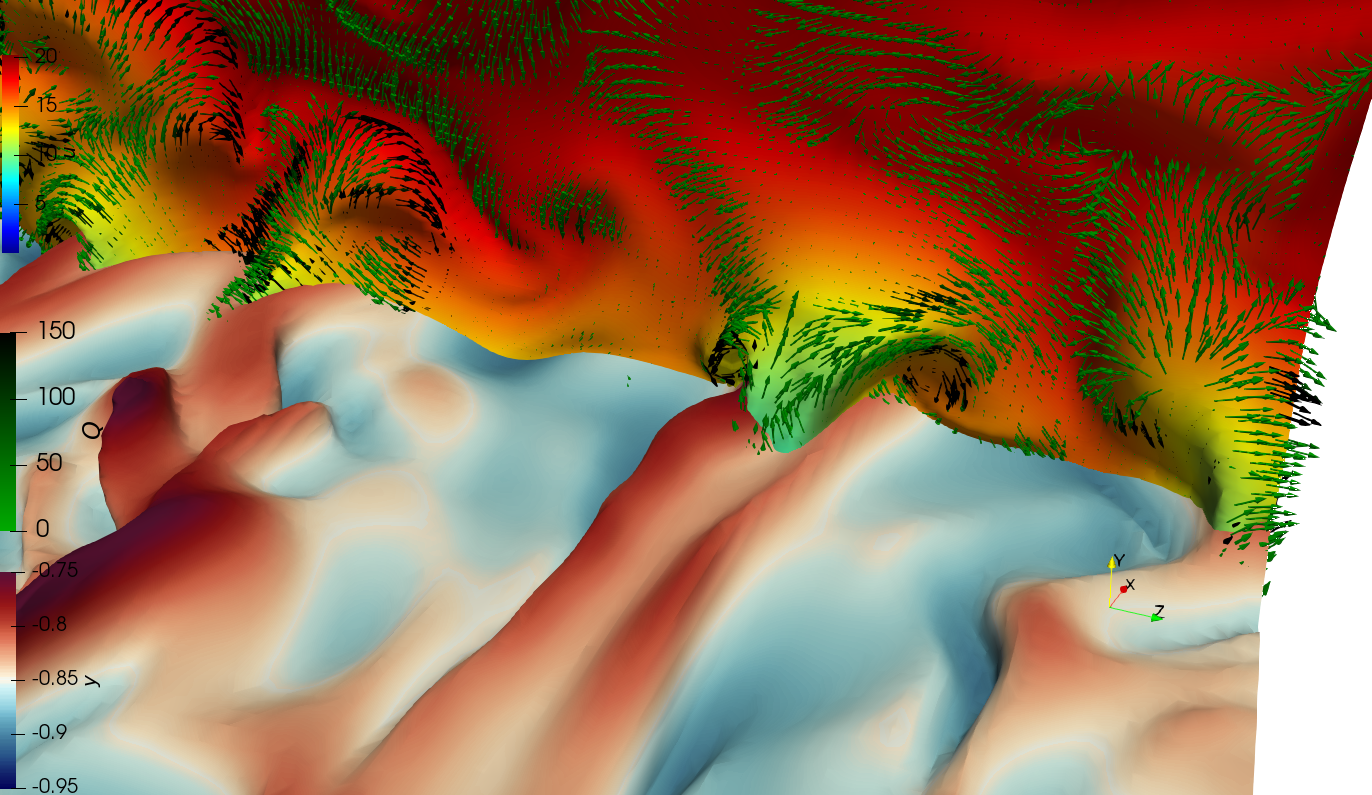}
    \end{minipage}
\caption{ Time-evolved material surfaces at $t^{+}=8$ corresponding to the initially constant-streamwise plane $x/\delta=1.5$ and the initially constant-wall-normal plane $y/\delta=-0.85$ ($y^{+}=30$) at $t^{+}=0$. 
The streamwise material surface is colored by the instantaneous 
streamwise velocity, superimposed with the velocity vectors (colored by $Q$). The wall-normal material surface is colored with the wall-normal coordinate ($y/\delta$). 
}\label{fig:TCF_tracking_bothplanes}
\end{figure}

Using the procedure described in section \ref{sec:theory_LCS}, the backward-time FTLE field is  evaluated at $t^{+} = 12 + t_0^{+}$. At this time instance, figure \ref{fig:TCF_LCS} \color{black} ($a-d$) \color{black} compare the backward-time FTLE with the Q-criterion by visualizing a couple of hairpin structures at the wall. 
As the potential attracting LCSs are identified with the ridges (local maxima) of the backward-time FTLE, 
they are commonly 
visualized through two-dimesional planes rather than three-dimensional iso-surfaces \citep{green2007}. Nonetheless, we first look at iso-surfaces of the backward-time FTLE to get a global picture of the FTLE field. 
Figure \ref{fig:TCF_LCS}$(a)$ shows that the iso-surfaces of \color{black}$\Lambda^{t_0}_{t} \delta/u_{\tau}=20$ \color{black} approximately coincide with the hairpin vortices identified by iso-surface of the Q-criterion. 
However,  a single FTLE iso-surface does not faithfully depict the boundaries of Lagrangian structures, and it is more appropriate to visualize a range of constant-FTLE surfaces (or volume) above a prescribed high value as in \cite{green2007}. Conversely, the negative volume etched by plotting constant-FTLE surfaces \textit{below} a prescribed FTLE limit can be visualized and compared with the Q-criterion structures. Figure \ref{fig:TCF_LCS}$(b)$ utilizes this visualization technique by plotting the volume of the region with \color{black}$\Lambda^{t_0}_{t} \delta/u_{\tau} \leq 15$\color{black}. It is observed 
(from multiple view angles although not all shown here) 
that the hairpin vortices from the Q-criterion fill up the negative region created by this volume, showing that the boundaries of coherent structures can indeed be delineated using FTLE. In figure \ref{fig:TCF_LCS}$(c)$ and $(d)$, FTLE is plotted in a streamwise and spanwise cut-plane, respectively, to illustrate the two-dimensional skeleton of LCS. It is clear that the regions of high FTLE closely correlate with the vortical structures given by the iso-surfaces of \color{black} the $Q$-criterion \color{black}. The ``mushroom" structure in figure \ref{fig:TCF_LCS}$(e)$, and the roll-up structure in figure \ref{fig:TCF_LCS}$(f)$ are indeed reminiscent of LCS
computed from a particle approach in \cite{green2007} for the same flow, as shown in the insets of these figures. Further evidence of the correlation between the FTLE and \color{black} $Q$-criterion \color{black} structures is provided by figure \ref{fig:TCF_LCS}$(g)$, 
\color{black}
 where the vortices indicated by the swirling velocity vectors (also 
corresponding to the regions of high $Q$) tend to be 
 bounded by regions of locally high $\Lambda^{t}_{t_0}$, 
consistent with the finding of \cite{green2007}.
\color{black}
Specifically, for the ``mushroom" structure in figure \ref{fig:TCF_LCS}$(g)$, the head of the mushroom closely wraps around the pair of counter-rotating vortices corresponding to the legs of the hairpin structure. Closer to the wall (to the bottom left and right of the mushroom), we observe more counter-rotating vortices likely related to the formation of low-speed streaks. This will be further evidenced through the visualization of material surfaces in figure \ref{fig:TCF_tracking_bothplanes}.

\begin{figure}
    \captionsetup{width=\linewidth}
    \begin{minipage}{0.32\textwidth}
    \includegraphics[trim=80 0 140 20,clip,width=\linewidth]{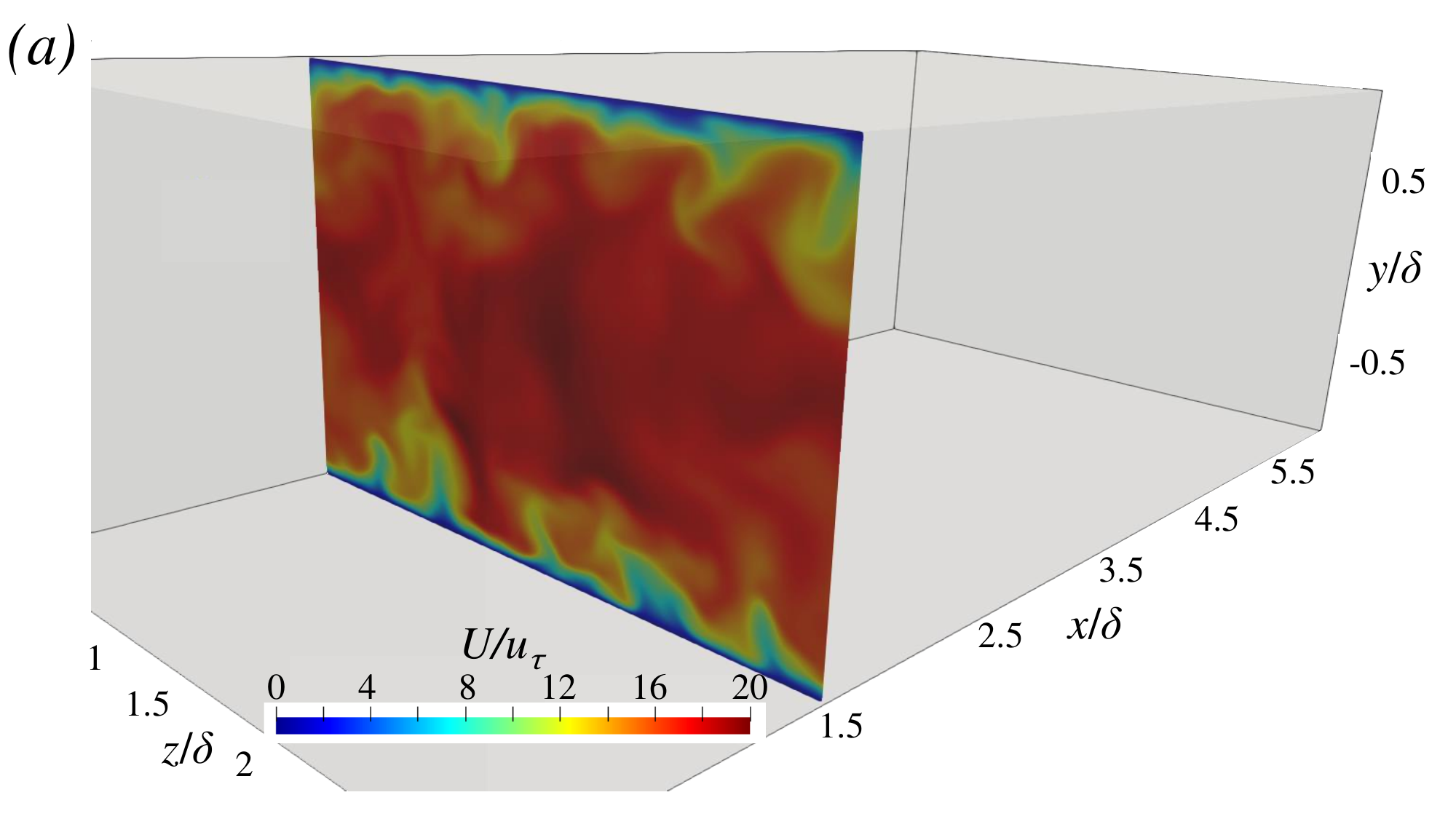}
    \end{minipage}
    \hspace{\fill} 
    \begin{minipage}{0.32\textwidth}
    \includegraphics[trim=80 0 140 20,clip,width=\linewidth]{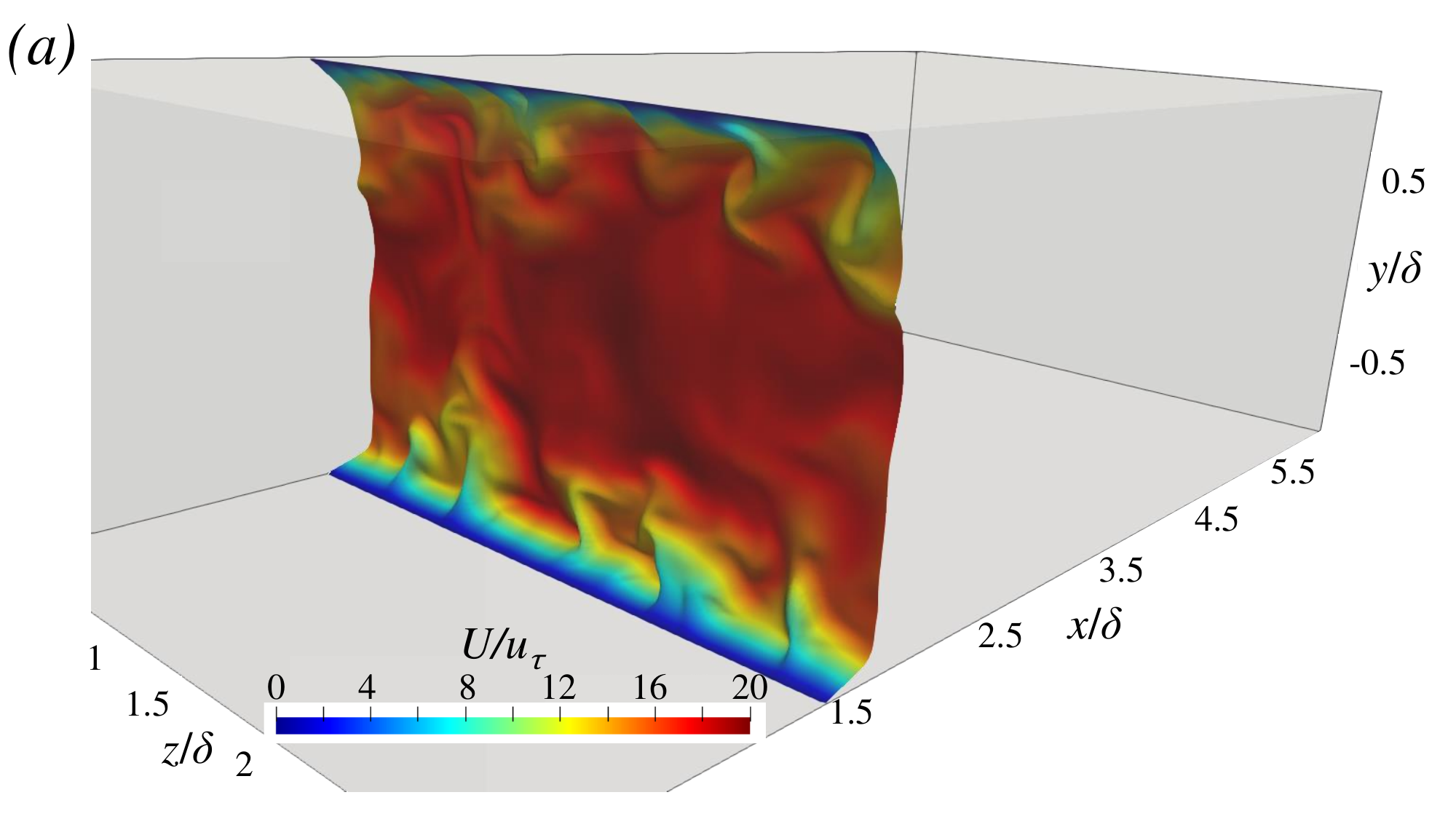}
    \end{minipage}
    \hspace{\fill} 
    \begin{minipage}{0.32\textwidth}
    \includegraphics[trim=80 0 140 20,clip,width=\linewidth]{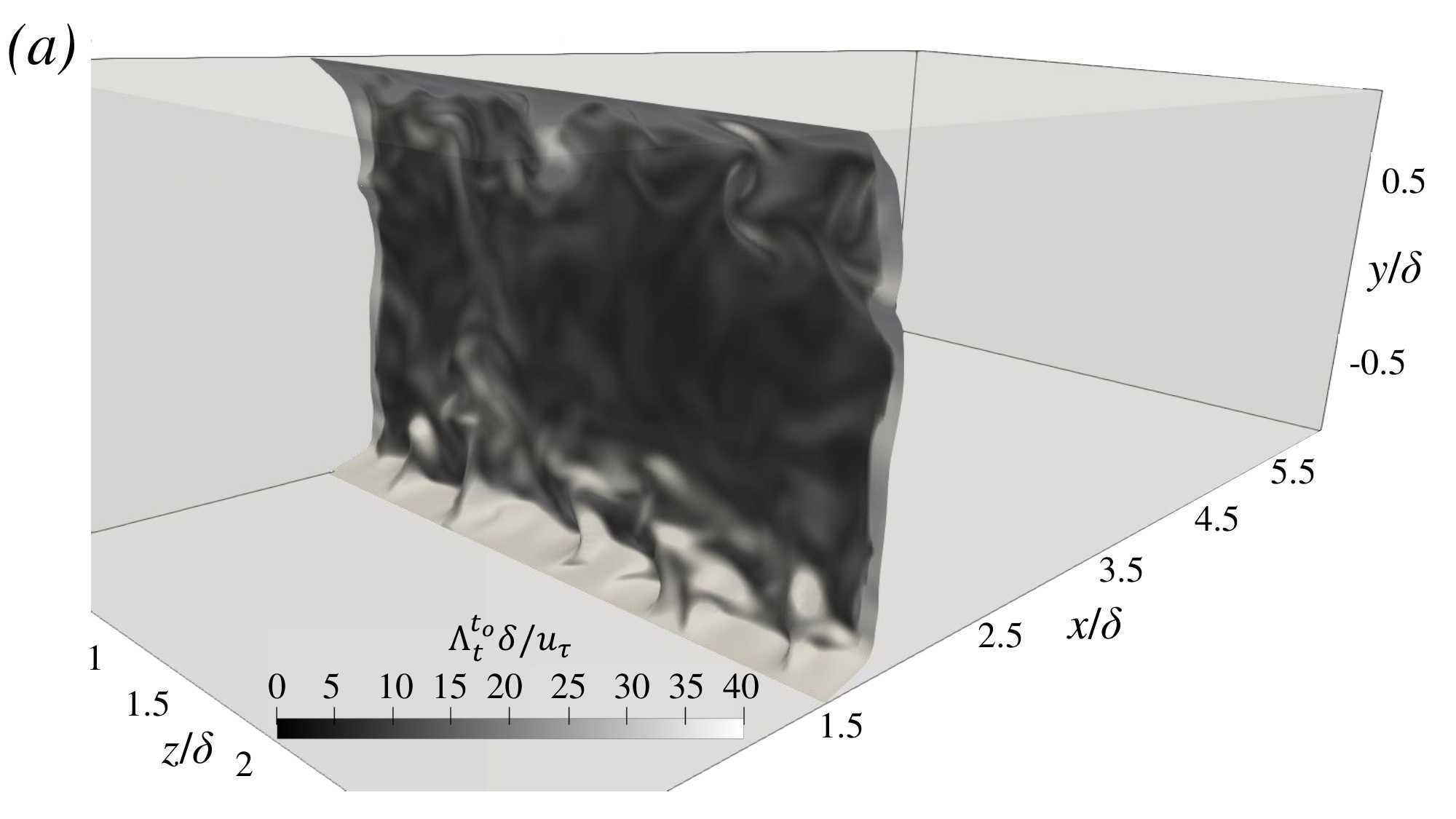}
    \end{minipage}
    \caption{Time evolution of the material surface corresponding to the constant-streamwise plane $x/\delta=1.5$ at $t^{+}=0$. The surface is evolved from $t^{+}=0$ (left most) to $t^{+}=4$ (middle), colored by the instantaneous streamwise velocity. The right most plot shows the same material surface at $t^{+}=4$ colored by the backward-time FTLE.}\label{fig:TCF_xplanetracking}
\end{figure}

A rather rudimentary but insightful way to visualize fluid element deformation  at a macroscopic level 
is  through  the evolution of 
initially planar material surfaces \citep{yeung2002lagrangian}, 
which is provided 
 directly by the reference map $\boldsymbol{\xi}$. Specifically, we track how the tracers that form a constant-streamwise and a constant-wall-normal plane at $t=0$ evolve in response to the background turbulent flow. Note that this is essentially the 3D counterpart of material-line tracking in figure \ref{fig:2D_TG_material_line}. Figure \ref{fig:TCF_xplanetracking} shows the time evolution of the plane identified by $\xi_1(x,y,z,t=0) = 1.5\delta$ from $t^{+}=0$ to $t^{+}=4$. It is interesting to note that the structures visible from the instantaneous velocity at $t^{+}=0$ leave a mark on the material surface as it evolves in time. Specifically, a mushroom structure indicative of a hairpin vortex is observed toward the top right of the plane. The regions of lower velocity within this structure lag behind the higher-velocity fluid particles outside it, leaving an imprint of the background turbulent flow structures on the material surface at $t^{+}=4$. In the right panel of figure \ref{fig:TCF_xplanetracking}, the backward-time FTLE is plotted on the deformed $\xi_1(x,y,z,t)$ surface showing that the imprinted structures correspond to regions of high FTLE.

\begin{figure}
    \captionsetup{width=\linewidth}
    \begin{minipage}{0.32\textwidth}
    \includegraphics[trim=160 10 70 10,clip,width=\linewidth]{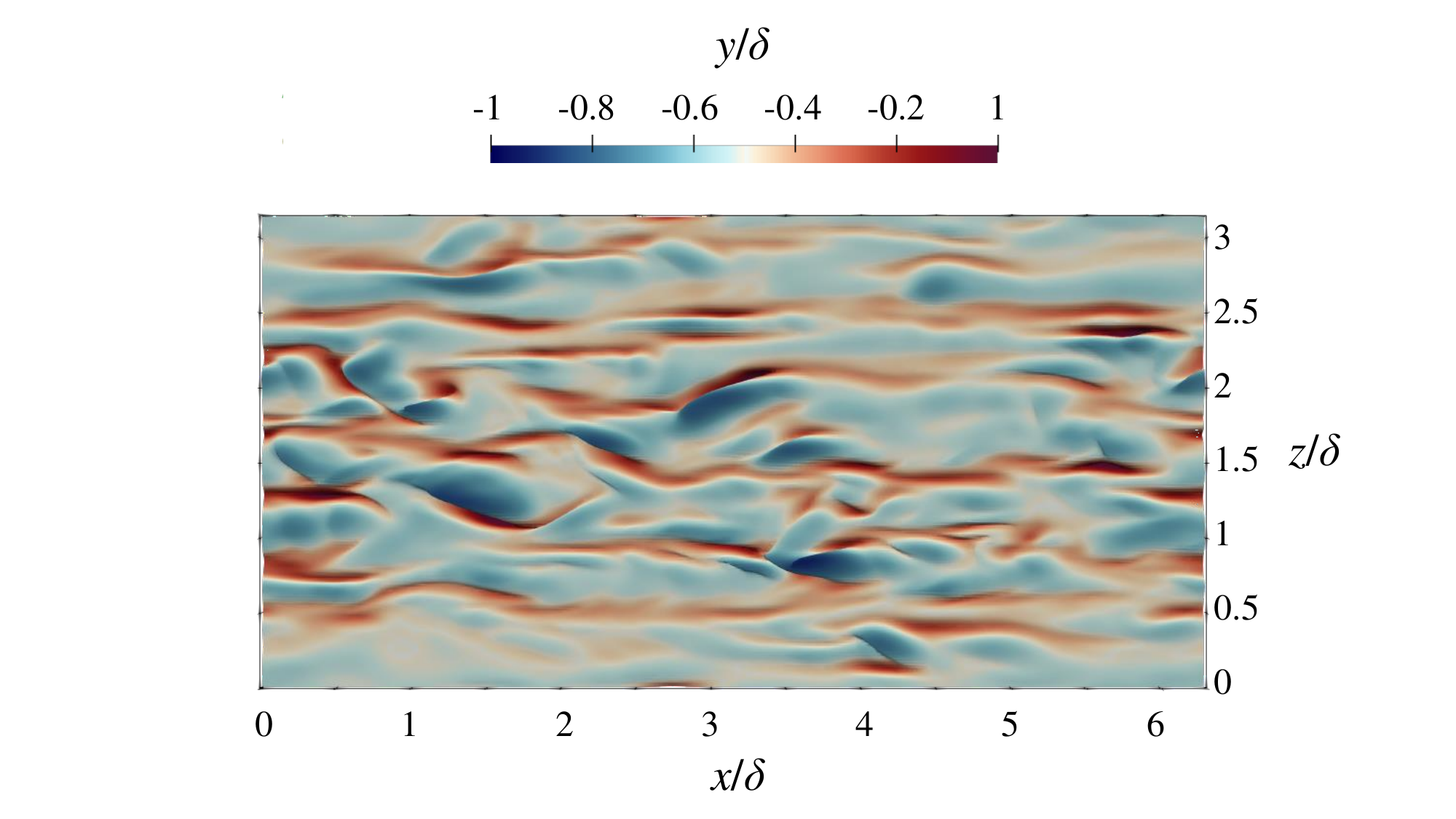}
    \end{minipage}
    \hspace{\fill} 
    \begin{minipage}{0.32\textwidth}
    \includegraphics[trim=160 10 70 10,clip,width=\linewidth]{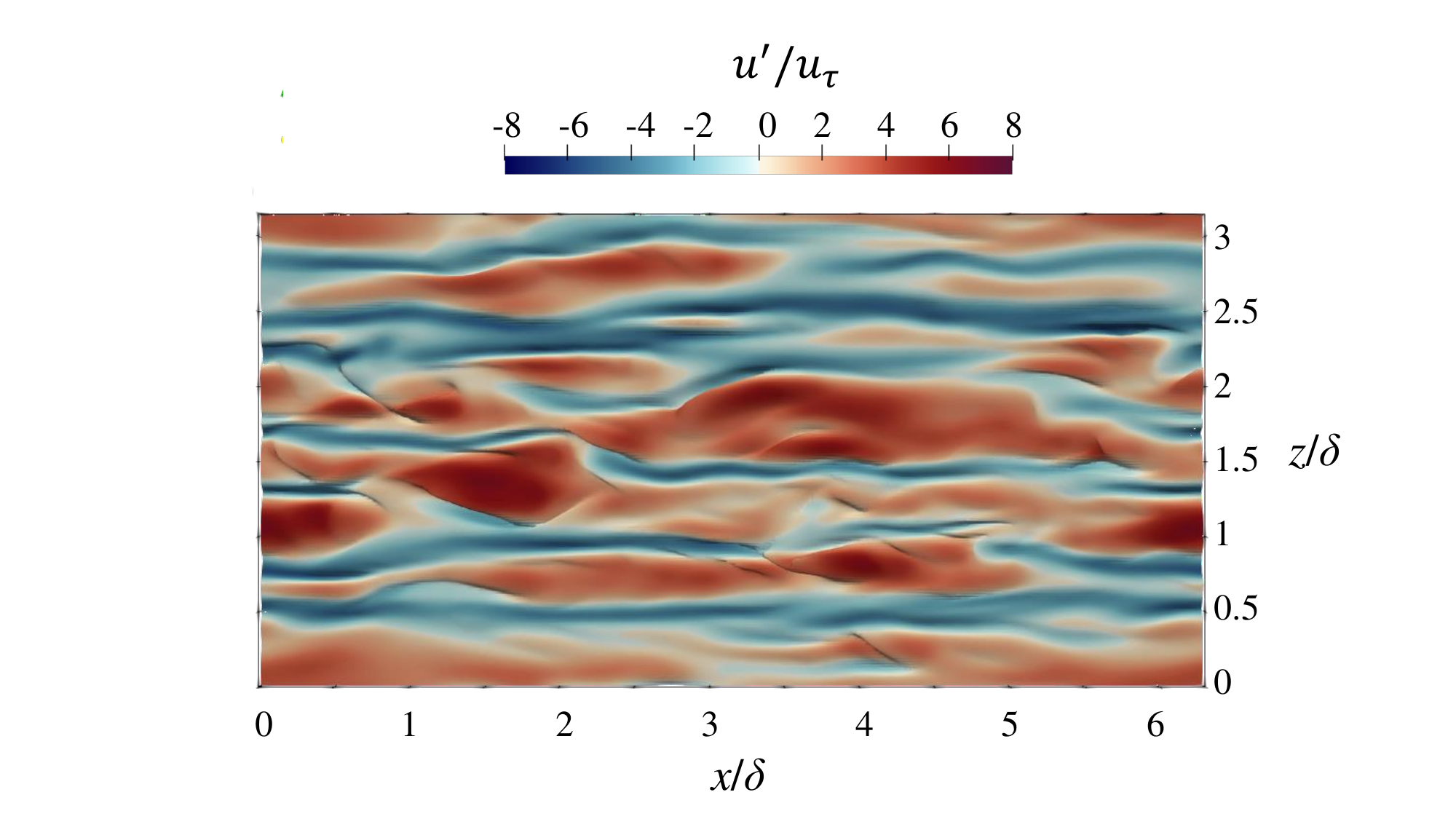}
    \end{minipage}
    \hspace{\fill} 
    \begin{minipage}{0.32\textwidth}
    \includegraphics[trim=160 10 70 10,clip,width=\linewidth]{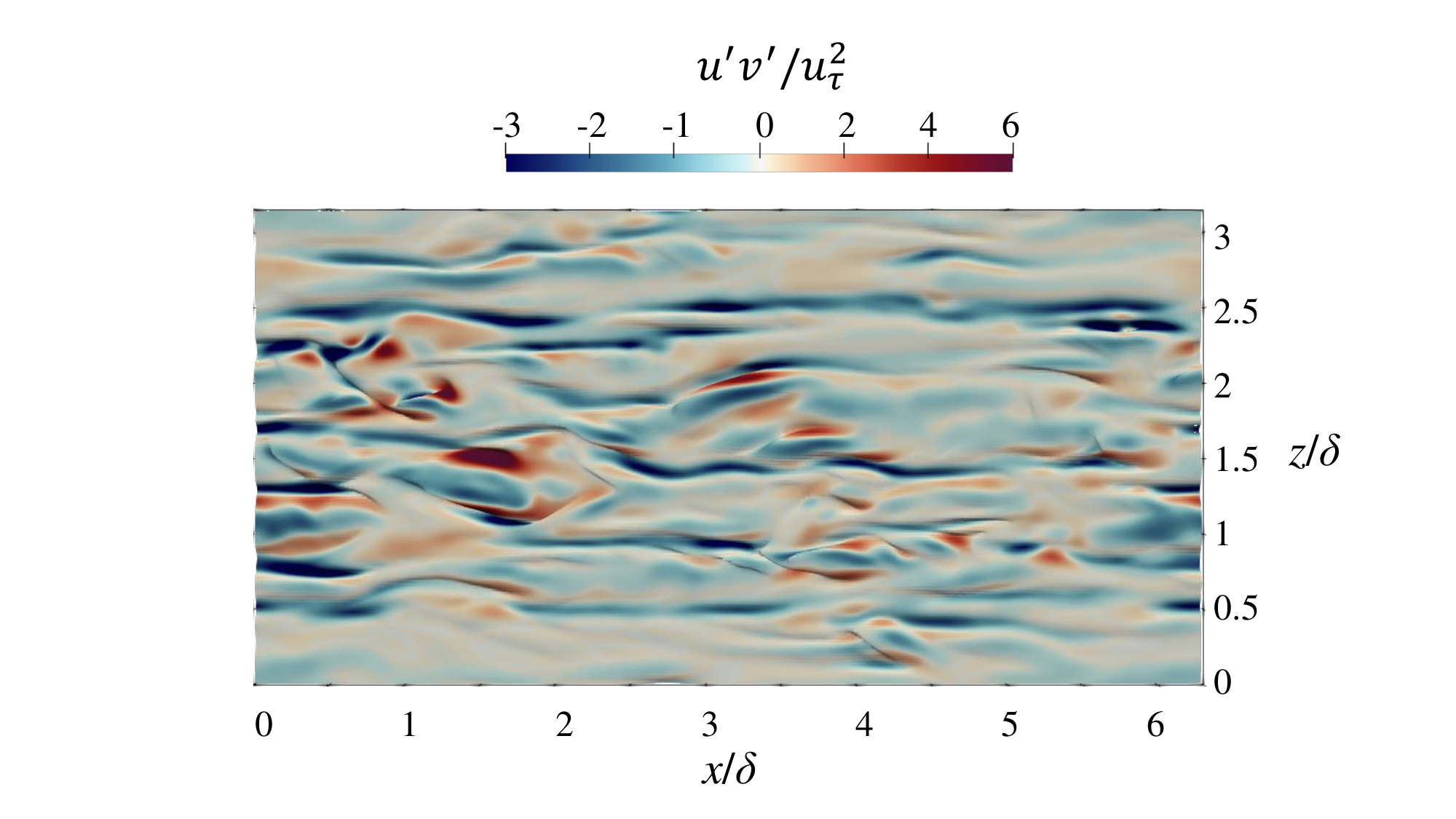}
    \end{minipage} 
    \begin{minipage}{0.32\textwidth}
    \includegraphics[trim=160 10 70 10,clip,width=\linewidth]{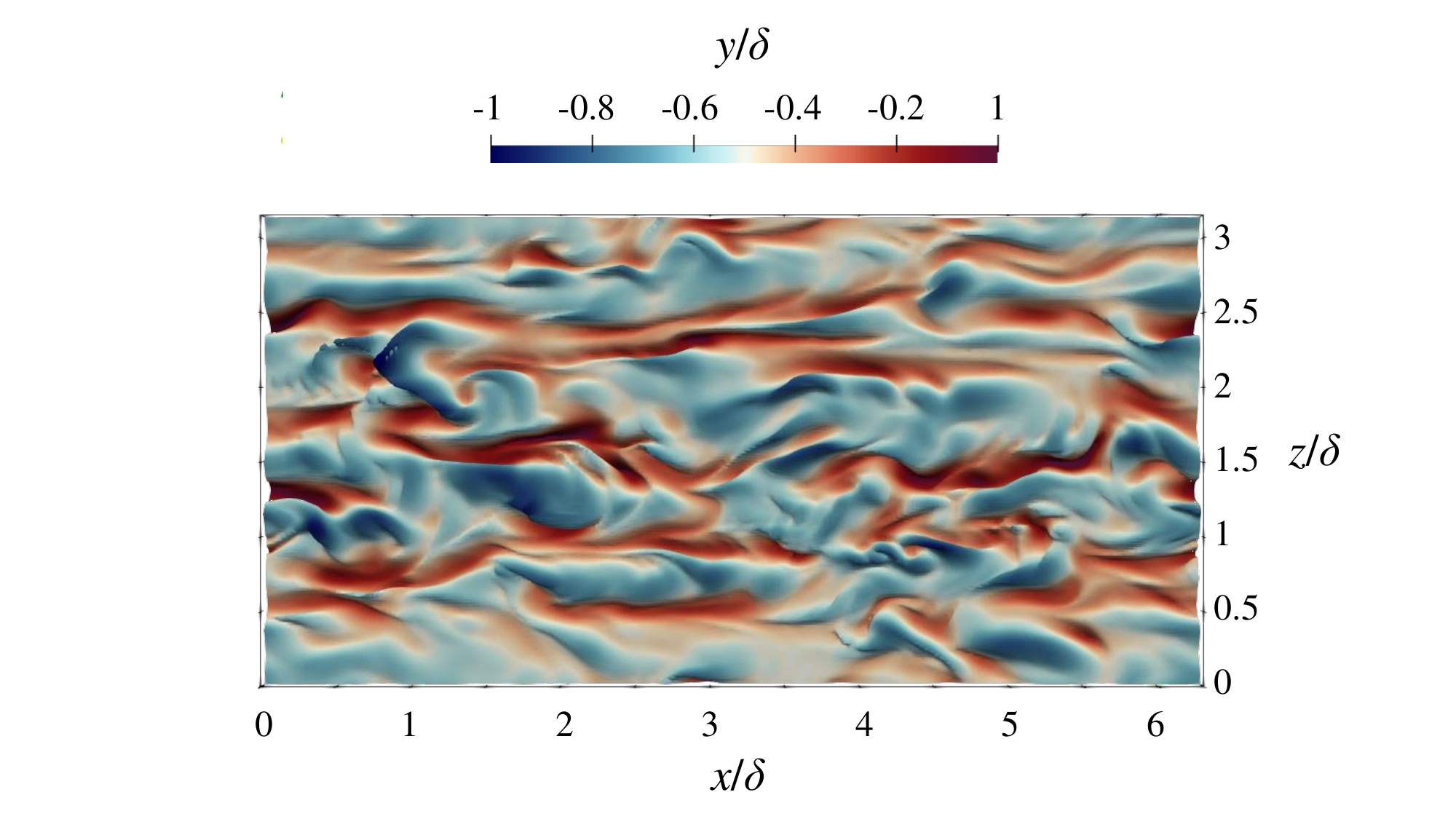}
    \end{minipage}
    \hspace{\fill} 
    \begin{minipage}{0.32\textwidth}
    \includegraphics[trim=160 10 70 10,clip,width=\linewidth]{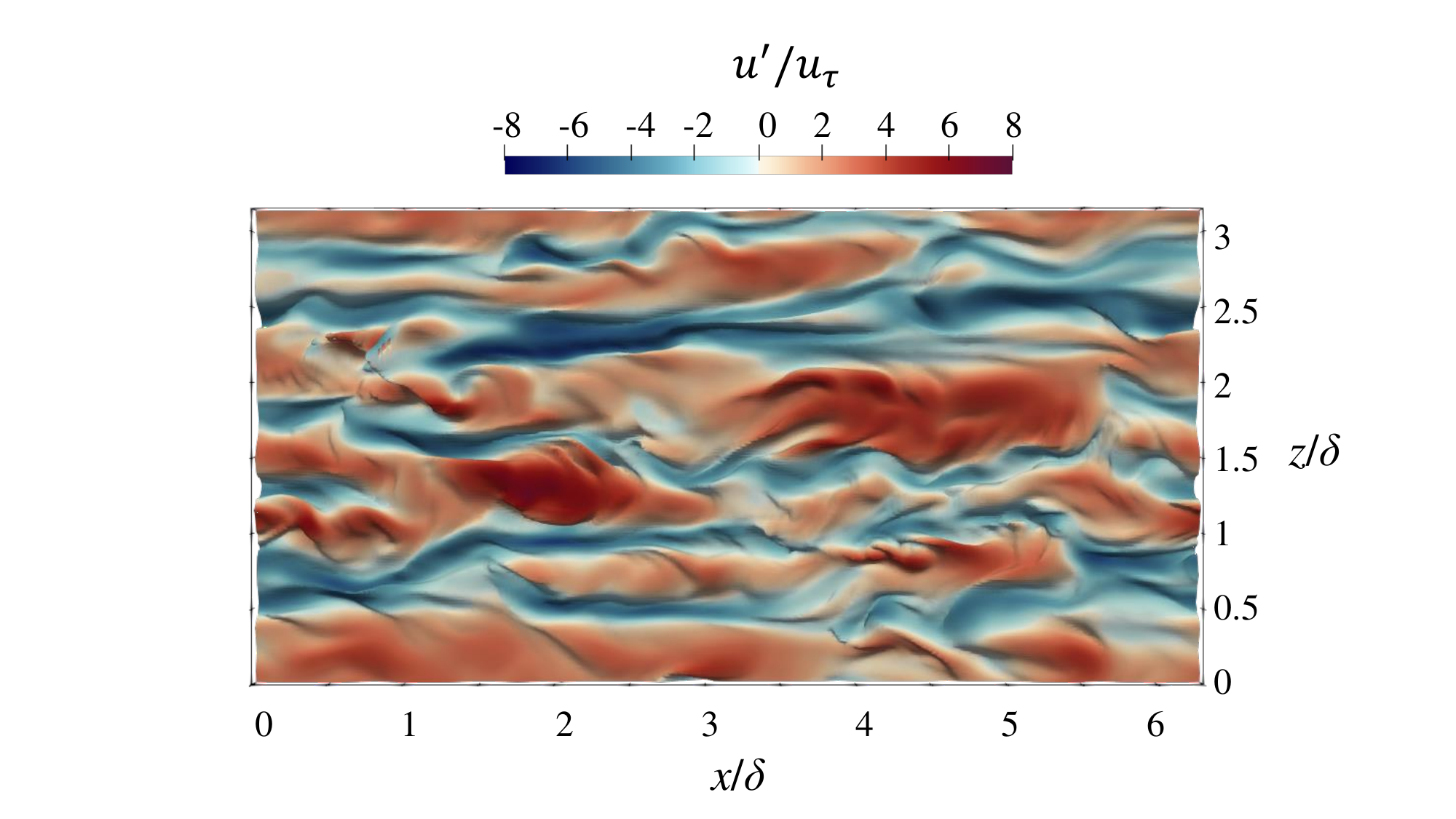}
    \end{minipage}
    \hspace{\fill} 
    \begin{minipage}{0.32\textwidth}
    \includegraphics[trim=160 10 70 10,clip,width=\linewidth]{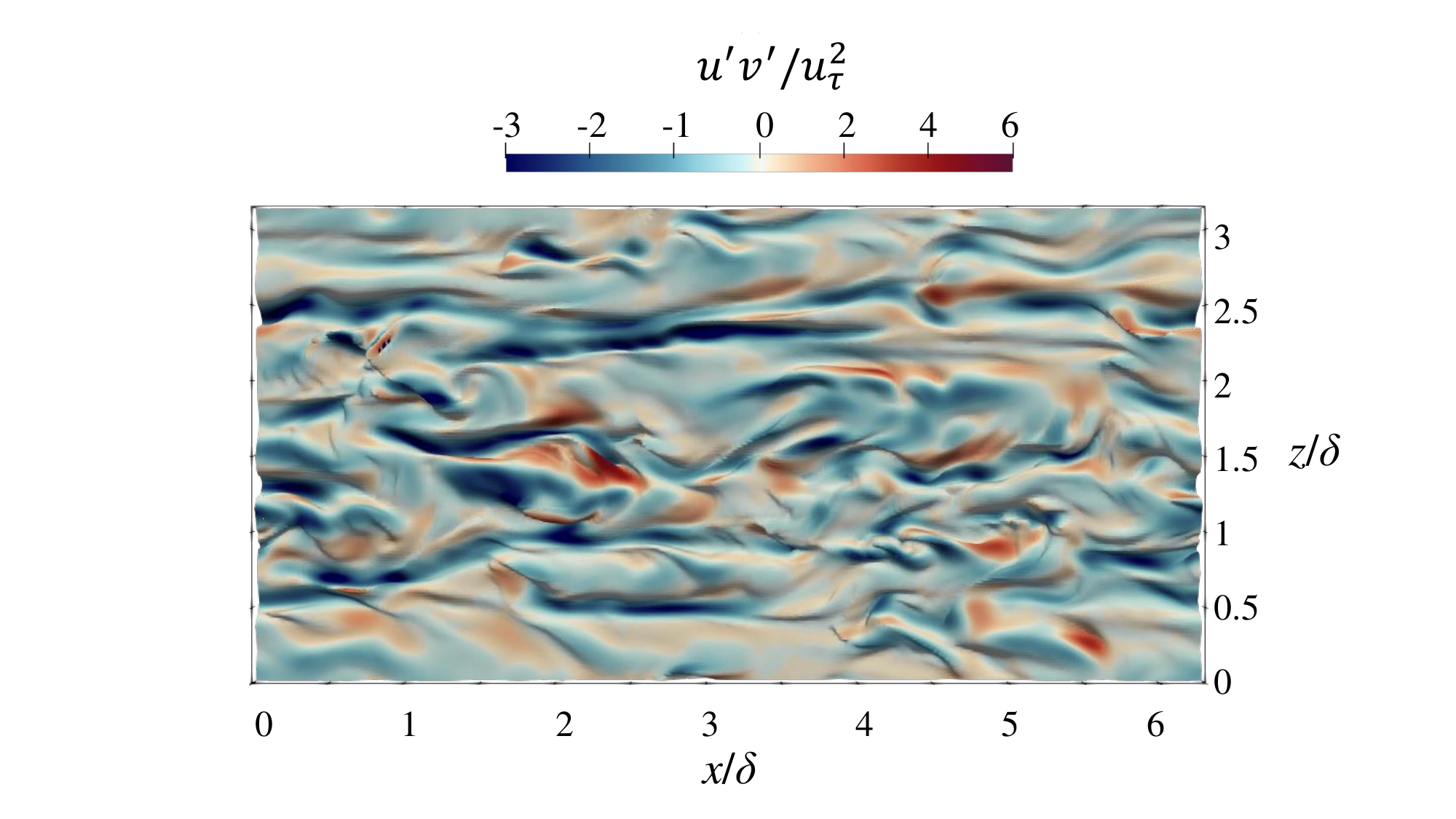}
    \end{minipage}
    \begin{minipage}{0.32\textwidth}
    \includegraphics[trim=160 10 70 10,clip,width=\linewidth]{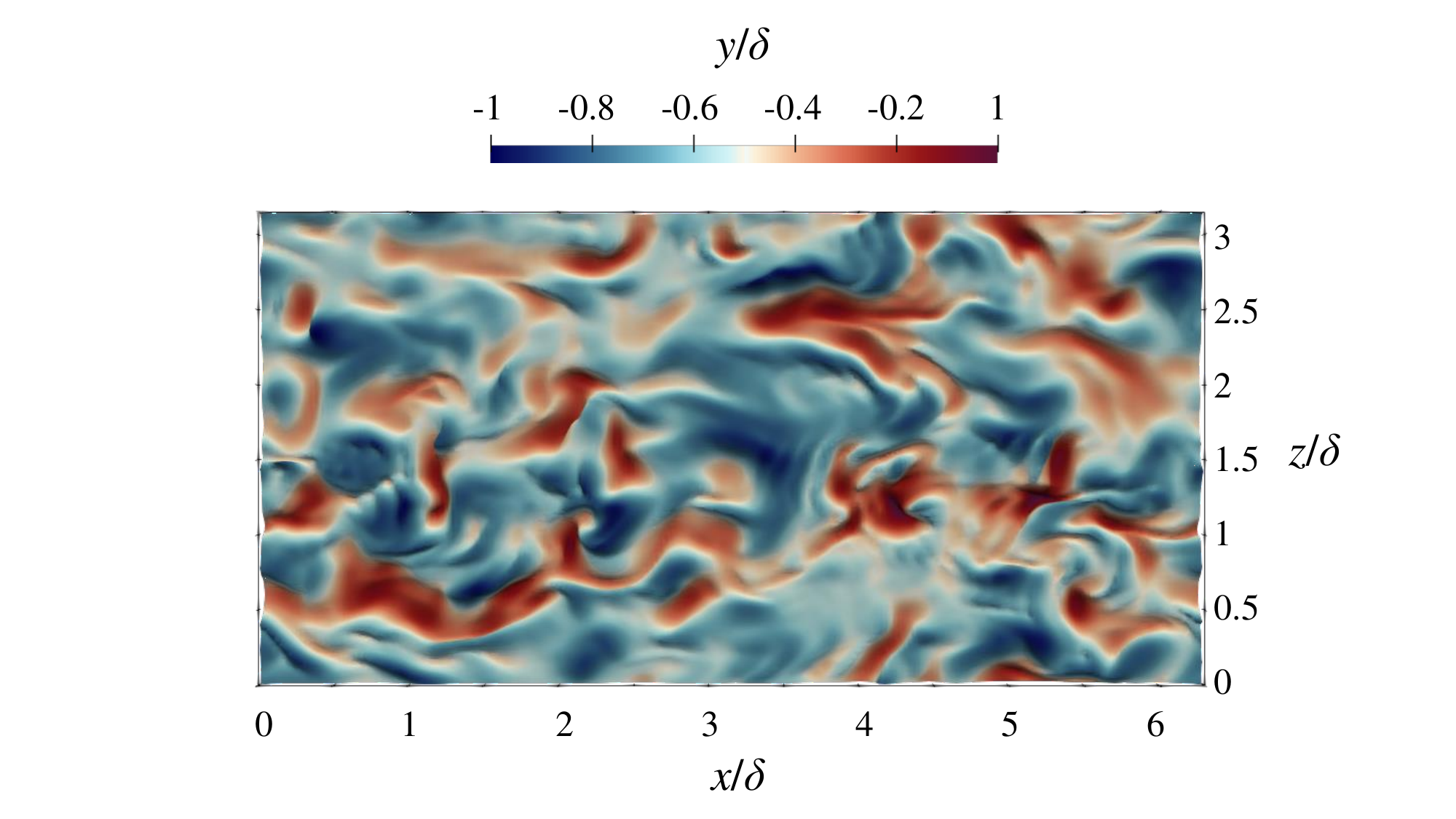}
    \end{minipage}
    \hspace{\fill} 
    \begin{minipage}{0.32\textwidth}
    \includegraphics[trim=160 10 70 10,clip,width=\linewidth]{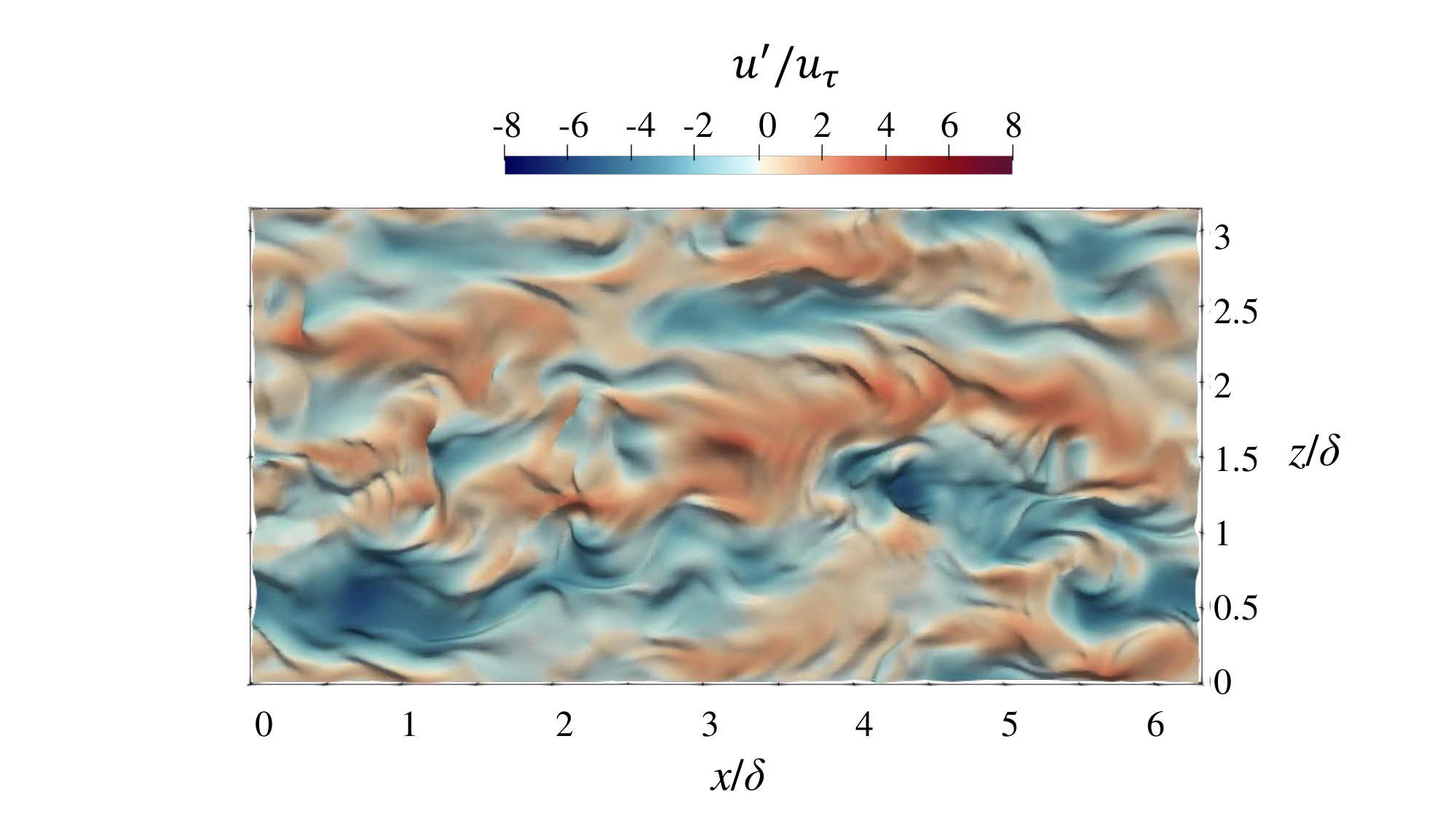}
    \end{minipage}
    \hspace{\fill} 
    \begin{minipage}{0.32\textwidth}
    \includegraphics[trim=160 10 70 10,clip,width=\linewidth]{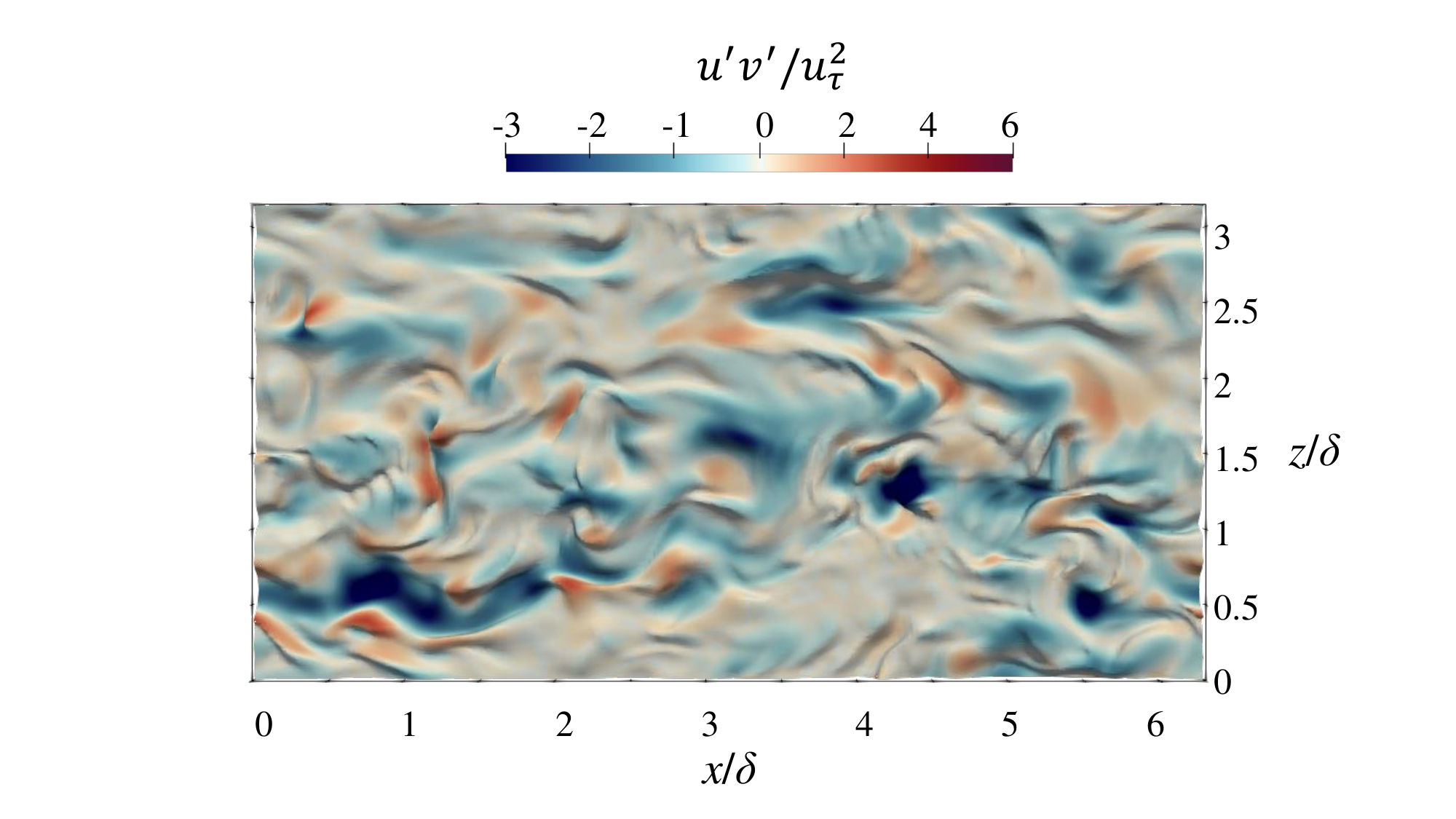}
    \end{minipage}    
\caption{Identification of ejection and sweep events through the time-evolved material surfaces of constant $\xi_2(x,y,z,t=0)$ at $t^{+}=8$. Each row corresponds to a different wall-normal plane from which $\xi_2(x,y,z,t=0)$ is initialized: first row, $y^{+}=10$; second row, $y^{+}=30$; third row, $y^{+}=70$. 
The first, second, and third columns are colored by 
$y/\delta$, $u'^+$, and $u'^+v'^+$, respectively. 
}\label{fig:TCF_yplanetracking}
\end{figure}

Figure \ref{fig:TCF_yplanetracking} shows three material surfaces at $t^{+}=8$ identified by the isosurfaces of $\xi_2$. Note that at $t=0$ these three material surfaces coincide with constant-wall-normal planes at $y/\delta=-0.95$ ($y^{+}=10$), $-0.85$ ($y^{+}=30$), and $-0.61$ ($y^{+}=70$). These locations approximately lie in the viscous sublayer, buffer layer, and log layer, respectively. The tracking of a material surface initialized at a constant-wall-normal distance is particularly helpful in identifying flow structures that cause significant vertical motion of the fluid particles such as sweep and ejection events. The left-most column in figure \ref{fig:TCF_yplanetracking} shows that the wall-normal deformation of the material surfaces gets stronger with increasing distance from the wall, which is perhaps an indication of the increasing eddy sizes away from the wall as per the attached-eddy hypothesis. The middle column in figure \ref{fig:TCF_yplanetracking} shows the same 
deformed surfaces now colored with the streamwise velocity fluctuations. It is observed that the elevations and depressions 
of the material surfaces 
(in the left column) tend to correspond respectively with the regions of negative and positive $u'$ (in the middle column), indicating the presence of ejection and sweep events. This becomes evident in the right-most column where $u'v'$ is plotted onto the deformed material surface. The strongly deformed regions (in the left column) are seen to correspond to regions of large negative $u'v'$ values (in the right column). Furthermore, the regions of negative $u'v'$ dominate all three layers next to the wall, but the buffer layer is most active in terms of these sweep and ejection events.

To get a more holistic and consolidated picture of the flow structures identified individually from the $x$- and $y$-plane tracking, in figure \ref{fig:TCF_tracking_bothplanes} we superimpose at  $t^{+}=8$ the 
constant $\xi_1$ and constant $\xi_2$ surfaces 
initially at $x/\delta=2.4$ and $y/\delta=-0.85$, respectively. Furthermore, to show the relation of these structures to those from the $Q$-criterion, the velocity vectors colored by the $Q$-criterion are plotted onto the deformed $\xi_1$ surface. We focus on one particular instance of a structure shown in the box. Here, the $\xi_2$ surface indicates the presence of an elongated low-speed streak through elevation in the material surface. Consistent with the well-studied mechanism for the low-speed streaks, this streak lies exactly between a pair of counter-rotating vortices depicted by the velocity vectors and the regions of high $Q$. The velocity vectors clearly show the ejection of fluid away from the wall between the two vortices. Furthermore, the lagging region of $\xi_1$ surface (described above in relation to figure \ref{fig:TCF_xplanetracking}) also coincides with the low-speed streak and is located between the counter-rotating vortex pair. The $\xi_1$ material surface exhibits twisting which is consistent with the direction of each of the counter-rotating vortices.

\begin{figure}
    \centering
    \captionsetup{width=\linewidth}
    \includegraphics[width=\textwidth]{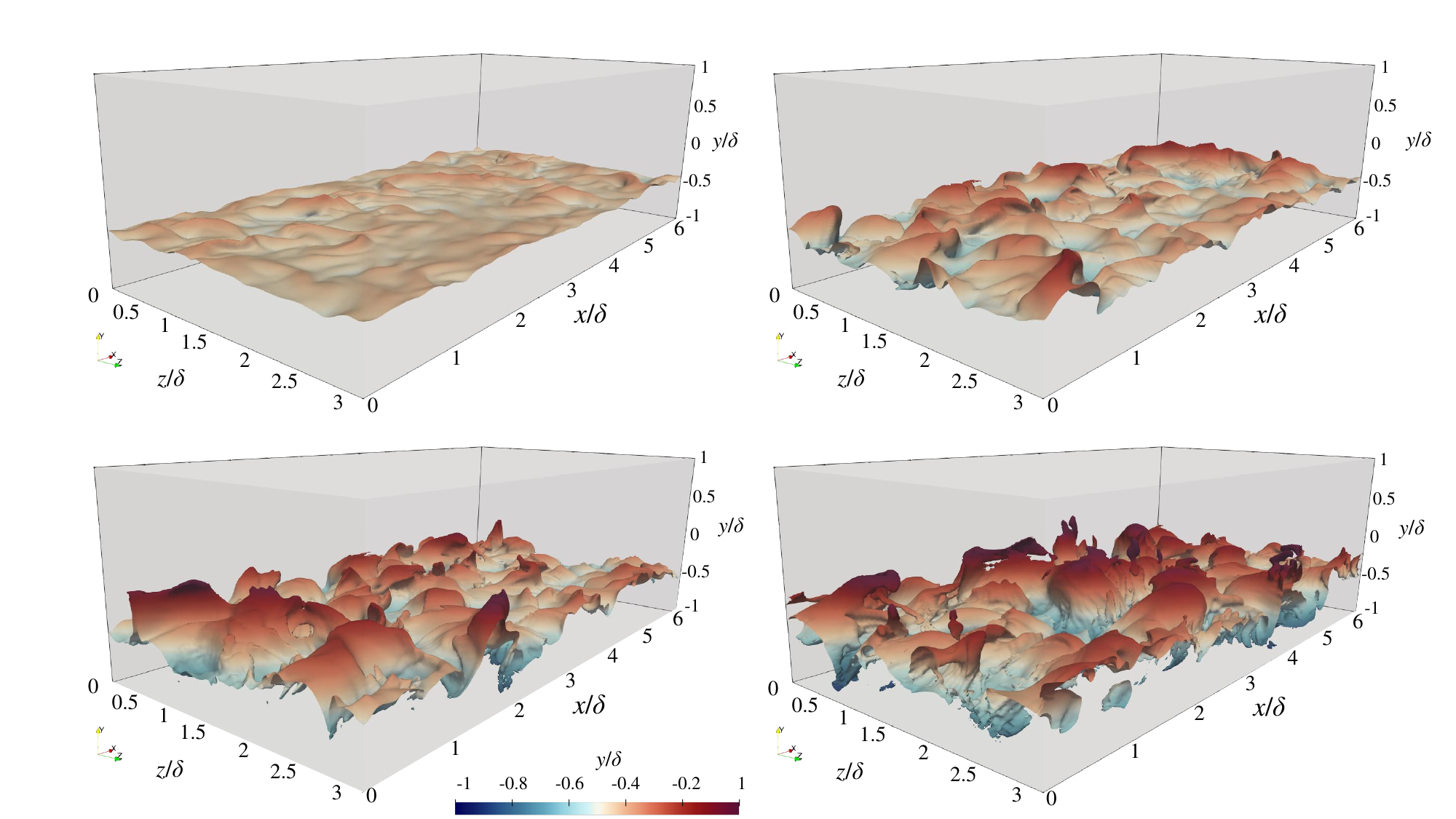}
    \caption{Time-evolution of the material surface initially corresponding to the constant-$y$ plane at $y^{+}=100$. Top left, $t^{+}=8$; top right, $t^{+}=20$, bottom left, $t^{+}=40$; bottom right, $t^{+}=60$.}
    \label{fig:TCF_yplanetracking_yplus100}
\end{figure}

In summary, the tracking of material surfaces over short durations of time presents a simple but natural way of identifying instantaneous Lagrangian structures. It is noted that material surfaces are found to retain the initial topologically connected surface for considerable times commensurate with local-flow time scales. For instance, the time scale of eddies in the logarithmic region according to \cite{lozano2019} is $\kappa y^+$ ($\kappa \approx 0.4$). Consistent with this, the material surface in figure \ref{fig:TCF_yplanetracking_yplus100} initialized at $y^{+}=100$ in the log layer remains topologically connected up to about $t^{+} = 40$, implying the current material surface tracking can be utilized usefully to study the Lagrangian dynamics of eddies therein.

\section{Discussion and Conclusion}


In this study, we have explored the application of the reference map technique, originally developed for the computation of solid stress in an Eulerian manner, to fluid mechanics within the context of Lagrangian kinematics. Traditional methods for computing Lagrangian kinematics involve the explicit tracking of tracer particles forward or backward in time, allowing for the calculation of the flow map and its gradient on an auxiliary background grid. The reference map offers an alternative by enabling the calculation of these quantities without the use of particles. This is accomplished through the Eulerian update of the advection equation governing the evolution in time of the spatial distribution of the reference map. The reference map records the take-off positions of tracers arriving at fixed mesh points, effectively tracking tracer particles backward in time, but in an implicit manner.

The reference map approach for Lagrangian kinematics offers enhanced flexibility and efficiency compared to the particle approach, particularly when the flow is derived through numerical simulation. The particle approach often necessitates seeding a substantial number of particles to reliably capture fluid particle trajectories across the flow domain seamlessly. The clustering of particles may result in the flow map (or its inverse) becoming a non-injective function, potentially leading to the ill-conditioning of the deformation gradient \citep{kafiabad2022grid}. Overall, the effective particle resolution surpasses that of the flow grid. When tracer trajectories are calculated on the fly with Navier-Stokes integration, this can result in a significant increase in computational cost, as well as additional overhead for large memory allocation. Achieving parallel load balancing of particles across processors can also pose challenges, especially when particle distribution in the processor space becomes highly irregular, as is often the case in inhomogeneous flows.
The choice of an interpolation scheme to approximate particle velocity at non-mesh points is recognized to significantly impact the clarity of the Lagrangian structure \citep{kafiabad2023computing}. High-order schemes, recommended to be at least fourth-order accurate and twice continuously differentiable (e.g., cubic spline \citep{yeung2002lagrangian}), play a crucial role in this regard. In contrast, the grid resolution for the reference map is inherently identical to that of the velocity and can be increased as needed. Load balancing is easily achieved with the standard domain decomposition already implemented for the flow solution. Velocity (or flux) data is consistently available at locations required by the reference map, eliminating the necessity for interpolation. The cost of the FTLE field calculation appears to be competitively comparable to the particle-based approach, as demonstrated in figure~\ref{fig:cost}.

The information implicitly embedded in the reference map regarding fluid particle trajectories provides a means to calculate measures of Lagrangian fluid element deformation solely from Eulerian data. These measures encompass displacement, deformation gradient tensor, and (left/right) Cauchy-Green tensor. They can be further leveraged to compute the backward-time Finite-Time Lyapunov Exponent (FTLE) field, with its ridges identified as candidates for the attracting LCS. We demonstrated the accuracy of FTLE calculations based on the reference map in comparison to the standard particle-based approach using a simple two-dimensional flow (2D Taylor-Green vortex). LCS detection utilizing the reference map was then applied to turbulent channel flow at $Re_\tau = 180$. The computed LCSs exhibited consistency with structures identified through the $Q$-criterion, where the FTLE ridges closely approximated the boundaries of the \color{black} $Q$-criterion \color{black} structures, as illustrated in \cite{green2007}.
A notable drawback of the current approach, in contrast to widely used Eulerian vortex identification schemes, is the high cost associated with incorporating an additional three advection equations for the reference map. While Eulerian schemes reveal coherent structures based on the instantaneous velocity field, LCS based on the concept of finite-time dynamical systems necessitates fluid particle trajectories over a specified time interval. Given that dominant LCSs tend to converge over relatively short time scales \citep{huang2022lagrangian, green2007}, this increased computational cost may be confined to a small fraction of the total simulation cost by integrating the reference map equation only during the time it is needed.

While the reference map does not explicitly track a particle with a fixed label, its ability to return the take-off positions of tracers arriving at grid points makes it well-suited for material surface tracking. Various shapes of material surfaces at the initial time can be realized through a judicious choice of the functional of the reference map. We have demonstrated the evolution of material surfaces with initial conditions set as streamwise-constant or wall-normal-constant planes in turbulent channel flow. This capability is notably unique to the reference map, as the reconstruction and visualization of equivalent surfaces with purely Lagrangian approaches are not straightforward.
The time evolution of material surfaces provides valuable insights into the Lagrangian landscape of turbulent momentum transport, which remains obscured in the Eulerian velocity field. Figure \ref{fig:TCF_yplanetracking}, in a sense, can be considered as a visual Lagrangian quadrant analysis. Although it lacks quantitative information, it offers an immediate three-dimensional visual account of key events to be analyzed through quadrant analysis. These events include ejections and sweeps identified from the vertical displacement of material surfaces, the signs and intensities of associated velocity fluctuations ($u'$ and $v'$), the identification of low-/high-speed streaks, and confirmation of the dominance of Reynolds-stress generating events in the second and fourth quadrants.

In conclusion, this paper explores the application of the reference map technique to fluid flows, with a focus on the computation of LCS  and the tracking of material surfaces, all in an Eulerian manner. It provides a convenient method and a promising direction for future investigations into Lagrangian kinematics and dynamics of fluid flows. One immediate application of the reference map technique could be in characterizing the deformation of Lagrangian fluid elements. For example, the deformation gradient fully characterizes the deformation of infinitesimal material lines as well as oriented material areas (Nanson's formula) \citep{gurtin_mechanics_2010}.








\section*{Acknowledgement}
This research was sponsored by 
 a faculty startup grant and the Ashton fellowship 
 provided by  the University of Pennsylvania. 

\section*{Declaration of interests} The authors report no conflict of interest.

\appendix

\section{\label{app:duality} Duality  between the backward- and forward-time FTLEs: Proof of Eq.~(\ref{eq:ftle_duality}).}

The largest forward-time FTLE at position $\textbf{X}$ is defined as 
\begin{equation}
    \Lambda^{t}_{t_0}(\textbf{X}) = \frac{1}{|t - t_0|}\text{ln}\sqrt{\lambda_{\text{max}}(\textbf{C}^{t}_{t_0}(\textbf{X}))}, 
    \label{a:l_ftle}
\end{equation}
where $\textbf{C}^{t}_{t_0}(\textbf{X}) = (\textbf{F}^{t}_{t_0}(\textbf{X}))^T\textbf{F}^{t}_{t_0}(\textbf{X})$ is the right Cauchy-Green deformation tensor of the forward map. 
One can show the following from the chain rule, 
\begin{equation}
    \textbf{F}^{t_0}_{t}(\textbf{x})\textbf{F}^{t}_{t_0}(\textbf{X}) = \frac{\partial \textbf{X}}{\partial \textbf{x}}\frac{\partial \textbf{x}}{\partial \textbf{X}} = \textbf{I}, 
\end{equation}
and therefore $\textbf{F}^{t}_{t_0}(\textbf{X}) = (\textbf{F}^{t_0}_{t}(\textbf{x}))^{-1}$. Using this result on the right Cauchy-Green tensor, 
\begin{equation}
    \begin{split}
        \textbf{C}^{t}_{t_0}(\textbf{X}) 
        & = \left [(\textbf{F}^{t_0}_{t}(\textbf{x}))^{-1}\right ]^T(\textbf{F}^{t_0}_{t}(\textbf{x}))^{-1} \\
        & = \left [(\textbf{F}^{t_0}_{t}(\textbf{x}))^{T}\right ]^{-1}(\textbf{F}^{t_0}_{t}(\textbf{x}))^{-1} \\
        & = \left [\textbf{F}^{t_0}_{t}(\textbf{x})(\textbf{F}^{t_0}_{t}(\textbf{x}))^{T}\right ]^{-1} \\
        & = \left [\textbf{B}_{t}^{t_0}(\textbf{x}) \right]^{-1}, 
    \end{split}
\end{equation}
where $\textbf{B}_{t}^{t_0}(\textbf{x})$ is the left Cauchy-Green deformation tensor of the backward flow map. Thus, we have 
\begin{equation}
    \lambda_{\text{max}}(\textbf{C}^{t}_{t_0}(\textbf{X})) = \frac{1}{\lambda_{\text{min}}(\textbf{B}_{t}^{t_0}(\textbf{x}))}. 
    \label{a:lamda_max}
\end{equation}
Next, we show the eigenvalues of $\textbf{B}_{t}^{t_0}(\textbf{x})$ are equal to the eigenvalues of $\textbf{C}_{t}^{t_0}(\textbf{x})$. Let $\textbf{r}$ be an eigenvector of $\textbf{B}_{t}^{t_0}(\textbf{x})$ with corresponding eigenvalue $\lambda$, then
\begin{equation}
    \textbf{F}^{t_0}_{t}(\textbf{x})(\textbf{F}^{t_0}_{t}(\textbf{x}))^{T} \textbf{r} = \lambda \textbf{r}. 
\end{equation}
Pre-multiplying 
this relation
by $(\textbf{F}^{t_0}_{t}(\textbf{x}))^{T}$, 
\begin{equation}
    \textbf{C}_{t}^{t_0}(\textbf{x}) (\textbf{F}^{t_0}_{t}(\textbf{x}))^{T} \textbf{r} = \lambda (\textbf{F}^{t_0}_{t}(\textbf{x}))^{T}\textbf{r}. 
\end{equation}
Therefore, $\lambda$ is also an eigenvalue of $\textbf{C}_{t}^{t_0}(\textbf{x})$ with eigenvector $(\textbf{F}^{t_0}_{t}(\textbf{x}))^{T}\textbf{r}$. Therefore, $\lambda_{\text{min}}(\textbf{B}_{t}^{t_0}(\textbf{x})) = \lambda_{\text{min}}(\textbf{C}_{t}^{t_0}(\textbf{x}))$. 
Using this result in Eq.~(\ref{a:lamda_max}),
\begin{equation}
    \lambda_{\text{max}}(\textbf{C}^{t}_{t_0}(\textbf{X})) = \frac{1}{\lambda_{\text{min}}(\textbf{B}_{t}^{t_0}(\textbf{x}))} = \frac{1}{\lambda_{\text{min}}(\textbf{C}_{t}^{t_0}(\textbf{x}))} = \frac{1}{\lambda_{\text{min}}(\textbf{C}_{t}^{t_0}(\boldsymbol{\chi}^{t}_{t_0}(\textbf{X})))}. 
\end{equation}
Substituting this result into Eq.~(\ref{a:l_ftle}) produces 
the duality relation in Eq.~(\ref{eq:ftle_duality}). 
\begin{equation}
    \begin{split}
        \Lambda^{t}_{t_0}(\textbf{X}) 
        & = \frac{1}{|t - t_0|}\text{ln}\sqrt{\lambda_{\text{max}}(\textbf{C}^{t}_{t_0}(\textbf{X}))} \\
        & = \frac{1}{|t - t_0|}\text{ln}\left[ \lambda_{\text{min}}(\textbf{C}_{t}^{t_0}(\boldsymbol{\chi}^{t}_{t_0}(\textbf{X}))) \right]^{-1/2} \\
        & = -\frac{1}{|t - t_0|}\text{ln}\left[ \lambda_{\text{min}}(\textbf{C}_{t}^{t_0}(\boldsymbol{\chi}^{t}_{t_0}(\textbf{X}))) \right]^{1/2} \\
        & = - \Gamma^{t_0}_t(\boldsymbol{\chi}^{t}_{t_0}(\textbf{X})). 
    \end{split}
\end{equation}
The utility of the above equation is as follows. 
In the reference map approach, the smallest backward-time FTLE can be first computed in the current configuration (or the grid) 
($\Gamma^{t_0}_t(\textbf{x})$). 
The negative of this can be considered as the 
largest forward-time FTLE ($\Lambda^{t}_{t_0}$),  graphed not on \textbf{X} but 
on the forward-flow image of \textbf{X}. 
If $\Lambda^{t}_{t_0}$ needs to be graphed on 
\textbf{X}, $\textbf{X} = \boldsymbol{\xi}(\textbf{x}, t)$ 
can be used in principle to build the initial/reference configurations \textbf{X}.


\bibliographystyle{jfm}
\bibliography{jfm}

\providecommand{\noopsort}[1]{}\providecommand{\singleletter}[1]{#1}%
\begin{thebibliography}{63}
\expandafter\ifx\csname natexlab\endcsname\relax\def\natexlab#1{#1}\fi
\def\au#1{#1} \def\ed#1{#1} \def\yr#1{#1}\def\at#1{#1}\def\jt#1{\textit{#1}} \def\bt#1{#1}\def\bvol#1{\textbf{#1}} \def\vol#1{#1} \def\pg#1{#1} \def\publ#1{#1}\def\arxiv#1{#1}\def\org#1{#1}\def\st#1{\textit{#1}}

\bibitem[Bettencourt {\em et~al.\/}(2013)Bettencourt, L{\'o}pez \& Hern{\'a}ndez-Garc{\'\i}a]{bettencourt2013characterization}
{\sc \au{Bettencourt, Joao~H}, \au{L{\'o}pez, Crist{\'o}bal} \& \au{Hern{\'a}ndez-Garc{\'\i}a, Emilio}} \yr{2013}  \at{Characterization of coherent structures in three-dimensional turbulent flows using the finite-size lyapunov exponent}.  \jt{Journal of Physics A: Mathematical and Theoretical}  \bvol{46}~(25),  \pg{254022}.

\bibitem[Bouffanais {\em et~al.\/}(2007)Bouffanais, Deville \& Leriche]{bouffanais2007large}
{\sc \au{Bouffanais, Roland}, \au{Deville, Michel~O} \& \au{Leriche, Emmanuel}} \yr{2007}  \at{Large-eddy simulation of the flow in a lid-driven cubical cavity}.  \jt{Physics of Fluids}  \bvol{19}~(5).

\bibitem[Chakraborty {\em et~al.\/}(2005)Chakraborty, Balachandar \& Adrian]{chakraborty2005relationships}
{\sc \au{Chakraborty, Pinaki}, \au{Balachandar, Sivaramakrishnan} \& \au{Adrian, Ronald~J}} \yr{2005}  \at{On the relationships between local vortex identification schemes}.  \jt{Journal of fluid mechanics}  \bvol{535},  \pg{189--214}.

\bibitem[Chong {\em et~al.\/}(1998)Chong, Soria, Perry, Chacin, Cantwell \& Na]{chong1998turbulence}
{\sc \au{Chong, Min~S}, \au{Soria, Julio}, \au{Perry, AE}, \au{Chacin, J}, \au{Cantwell, BJ} \& \au{Na, Y}} \yr{1998}  \at{Turbulence structures of wall-bounded shear flows found using dns data}.  \jt{Journal of Fluid Mechanics}  \bvol{357},  \pg{225--247}.

\bibitem[Duguet {\em et~al.\/}(2012)Duguet, Schlatter, Henningson \& Eckhardt]{duguet2012self}
{\sc \au{Duguet, Yohann}, \au{Schlatter, Philipp}, \au{Henningson, Dan~S} \& \au{Eckhardt, Bruno}} \yr{2012}  \at{Self-sustained localized structures in a boundary-layer flow}.  \jt{Physical review letters}  \bvol{108}~(4),  \pg{044501}.

\bibitem[Farazmand \& Haller(2012)]{farazmand2012computing}
{\sc \au{Farazmand, Mohammad} \& \au{Haller, George}} \yr{2012}  \at{Computing lagrangian coherent structures from their variational theory}.  \jt{Chaos: An Interdisciplinary Journal of Nonlinear Science}  \bvol{22}~(1).

\bibitem[Farazmand \& Haller(2013)]{farazmand2013attracting}
{\sc \au{Farazmand, Mohammad} \& \au{Haller, George}} \yr{2013}  \at{Attracting and repelling lagrangian coherent structures from a single computation}.  \jt{Chaos: An Interdisciplinary Journal of Nonlinear Science}  \bvol{23}~(2).

\bibitem[Ghia {\em et~al.\/}(1982)Ghia, Ghia \& Shin]{ghia1982}
{\sc \au{Ghia, UKNG}, \au{Ghia, Kirti~N} \& \au{Shin, CT}} \yr{1982}  \at{High-{Re} solutions for incompressible flow using the navier-stokes equations and a multigrid method}.  \jt{Journal of computational physics}  \bvol{48}~(3),  \pg{387--411}.

\bibitem[Goc {\em et~al.\/}(2021)Goc, Lehmkuhl, Park, Bose \& Moin]{goc2021large}
{\sc \au{Goc, Konrad~A}, \au{Lehmkuhl, Oriol}, \au{Park, George~Ilhwan}, \au{Bose, Sanjeeb~T} \& \au{Moin, Parviz}} \yr{2021}  \at{Large eddy simulation of aircraft at affordable cost: a milestone in computational fluid dynamics}.  \jt{Flow}  \bvol{1},  \pg{E14}.

\bibitem[Green {\em et~al.\/}(2007)Green, Rowley \& Haller]{green2007}
{\sc \au{Green, Melissa~A}, \au{Rowley, Clarence~W} \& \au{Haller, George}} \yr{2007}  \at{Detection of lagrangian coherent structures in three-dimensional turbulence}.  \jt{Journal of Fluid Mechanics}  \bvol{572},  \pg{111--120}.

\bibitem[Gurtin {\em et~al.\/}(2010)Gurtin, Fried \& Anand]{gurtin_mechanics_2010}
{\sc \au{Gurtin, Morton~E.}, \au{Fried, Eliot} \& \au{Anand, Lallit}} \yr{2010} {\em The {Mechanics} and {Thermodynamics} of {Continua}\/}, 1st edn.  \publ{Cambridge University Press}.

\bibitem[Haller(2005)]{haller2005objective}
{\sc \au{Haller, George}} \yr{2005}  \at{An objective definition of a vortex}.  \jt{Journal of fluid mechanics}  \bvol{525},  \pg{1--26}.

\bibitem[Haller(2015)]{haller_lagrangian_2015}
{\sc \au{Haller, George}} \yr{2015}  \at{Lagrangian {Coherent} {Structures}}.  \jt{Annu. Rev. Fluid Mech.}  \bvol{47}~(1),  \pg{137--162}.

\bibitem[Haller \& Beron-Vera(2012)]{haller2012geodesic}
{\sc \au{Haller, George} \& \au{Beron-Vera, Francisco~J}} \yr{2012}  \at{Geodesic theory of transport barriers in two-dimensional flows}.  \jt{Physica D: Nonlinear Phenomena}  \bvol{241}~(20),  \pg{1680--1702}.

\bibitem[Haller \& Sapsis(2011)]{haller_lagrangian_2011}
{\sc \au{Haller, George} \& \au{Sapsis, Themistoklis}} \yr{2011}  \at{Lagrangian coherent structures and the smallest finite-time {Lyapunov} exponent}.  \jt{Chaos: An Interdisciplinary Journal of Nonlinear Science}  \bvol{21}~(2),  \pg{023115}.

\bibitem[Haller \& Yuan(2000)]{haller_lagrangian_2000}
{\sc \au{Haller, G.} \& \au{Yuan, G.}} \yr{2000}  \at{Lagrangian coherent structures and mixing in two-dimensional turbulence}.  \jt{Physica D: Nonlinear Phenomena}  \bvol{147}~(3-4),  \pg{352--370}.

\bibitem[Hayat \& Park(2023{\natexlab{{\em a\/}}})]{hayat2023efficient}
{\sc \au{Hayat, Imran} \& \au{Park, George~Ilhwan}} \yr{2023{\natexlab{{\em a\/}}}}  \at{Efficient spectral implementation of ode wall model and the extension of integral wall model to unstructured les solvers}.  \jt{Journal of Computational Physics}  \bvol{487},  \pg{112175}.

\bibitem[Hayat \& Park(2023{\natexlab{{\em b\/}}})]{hayat2023wall}
{\sc \au{Hayat, Imran} \& \au{Park, George~Ilhwan}} \yr{2023{\natexlab{{\em b\/}}}}  \at{Wall-modeled large-eddy simulation of turbulent boundary layer with spatially varying pressure gradients}.  \jt{AIAA Journal}  \pg{pp. 1--16}.

\bibitem[He {\em et~al.\/}(2016)He, Pan, Feng, Gao \& Wang]{he_pan_feng_gao_wang_2016}
{\sc \au{He, Guo-Sheng}, \au{Pan, Chong}, \au{Feng, Li-Hao}, \au{Gao, Qi} \& \au{Wang, Jin-Jun}} \yr{2016}  \at{Evolution of lagrangian coherent structures in a cylinder-wake disturbed flat plate boundary layer}.  \jt{Journal of Fluid Mechanics}  \bvol{792},  \pg{274–306}.

\bibitem[Hu {\em et~al.\/}(2023)Hu, Hayat \& Park]{hu2023wall}
{\sc \au{Hu, Xiaohan}, \au{Hayat, Imran} \& \au{Park, George~Ilhwan}} \yr{2023}  \at{Wall-modelled large-eddy simulation of three-dimensional turbulent boundary layer in a bent square duct}.  \jt{Journal of Fluid Mechanics}  \bvol{960},  \pg{A29}.

\bibitem[Huang {\em et~al.\/}(2022)Huang, Borthwick \& Lin]{huang2022lagrangian}
{\sc \au{Huang, Chenyang}, \au{Borthwick, Alistair~GL} \& \au{Lin, Zhiliang}} \yr{2022}  \at{Lagrangian coherent structures in flow past a backward-facing step}.  \jt{Journal of Fluid Mechanics}  \bvol{947},  \pg{A4}.

\bibitem[Hunt {\em et~al.\/}(1988)Hunt, Wray \& Moin]{hunt1988eddies}
{\sc \au{Hunt, Julian~CR}, \au{Wray, Alan~A} \& \au{Moin, Parviz}} \yr{1988}  \at{Eddies, streams, and convergence zones in turbulent flows}.  \jt{Studying turbulence using numerical simulation databases, 2. Proceedings of the 1988 summer program} .

\bibitem[Hussain(1986)]{hussain_1986}
{\sc \au{Hussain, A. K. M.~Fazle}} \yr{1986}  \at{Coherent structures and turbulence}.  \jt{Journal of Fluid Mechanics}  \bvol{173},  \pg{303–356}.

\bibitem[Jain {\em et~al.\/}(2019)Jain, Kamrin \& Mani]{jain_conservative_2019}
{\sc \au{Jain, Suhas~S.}, \au{Kamrin, Ken} \& \au{Mani, Ali}} \yr{2019}  \at{A conservative and non-dissipative {Eulerian} formulation for the simulation of soft solids in fluids}.  \jt{Journal of Computational Physics}  \bvol{399},  \pg{108922}.

\bibitem[Jeong \& Hussain(1995)]{jeong1995identification}
{\sc \au{Jeong, Jinhee} \& \au{Hussain, Fazle}} \yr{1995}  \at{On the identification of a vortex}.  \jt{Journal of fluid mechanics}  \bvol{285},  \pg{69--94}.

\bibitem[Kafiabad(2022)]{kafiabad2022grid}
{\sc \au{Kafiabad, Hossein~A}} \yr{2022}  \at{Grid-based calculation of the lagrangian mean}.  \jt{Journal of Fluid Mechanics}  \bvol{940},  \pg{A21}.

\bibitem[Kafiabad \& Vanneste(2023)]{kafiabad2023computing}
{\sc \au{Kafiabad, Hossein~A} \& \au{Vanneste, Jacques}} \yr{2023}  \at{Computing lagrangian means}.  \jt{Journal of Fluid Mechanics}  \bvol{960},  \pg{A36}.

\bibitem[Kamrin {\em et~al.\/}(2012)Kamrin, Rycroft \& Nave]{kamrin_reference_2012}
{\sc \au{Kamrin, Ken}, \au{Rycroft, Chris~H.} \& \au{Nave, Jean-Christophe}} \yr{2012}  \at{Reference map technique for finite-strain elasticity and fluid–solid interaction}.  \jt{Journal of the Mechanics and Physics of Solids}  \bvol{60}~(11),  \pg{1952--1969}.

\bibitem[Kasten {\em et~al.\/}(2010)Kasten, Petz, Hotz, Hege, Noack \& Tadmor]{kasten2010lagrangian}
{\sc \au{Kasten, Jens}, \au{Petz, Christoph}, \au{Hotz, Ingrid}, \au{Hege, Hans-Christian}, \au{Noack, Bernd~R} \& \au{Tadmor, Gilead}} \yr{2010}  \at{Lagrangian feature extraction of the cylinder wake}.  \jt{Physics of fluids}  \bvol{22}~(9).

\bibitem[Kim {\em et~al.\/}(1987)Kim, Moin \& Moser]{Kim1987}
{\sc \au{Kim, John}, \au{Moin, Parviz} \& \au{Moser, Robert}} \yr{1987}  \at{Turbulence statistics in fully developed channel flow at low reynolds number}.  \jt{Journal of fluid mechanics}  \bvol{177},  \pg{133--166}.

\bibitem[Kuhlmann \& Roman{\`o}(2019)]{kuhlmann2019lid}
{\sc \au{Kuhlmann, Hendrik~C} \& \au{Roman{\`o}, Francesco}} \yr{2019}  \at{The lid-driven cavity}.  \jt{Computational Modelling of Bifurcations and Instabilities in Fluid Dynamics}  \pg{pp. 233--309}.

\bibitem[Leung(2011)]{leung_eulerian_2011}
{\sc \au{Leung, Shingyu}} \yr{2011}  \at{An {Eulerian} approach for computing the finite time {Lyapunov} exponent}.  \jt{Journal of Computational Physics}  \bvol{230}~(9),  \pg{3500--3524}.

\bibitem[Leung(2013)]{leung2013backward}
{\sc \au{Leung, Shingyu}} \yr{2013}  \at{The backward phase flow method for the eulerian finite time lyapunov exponent computations}.  \jt{Chaos: An Interdisciplinary Journal of Nonlinear Science}  \bvol{23}~(4).

\bibitem[Lozano-Dur{\'a}n \& Bae(2019)]{lozano2019}
{\sc \au{Lozano-Dur{\'a}n, Adri{\'a}n} \& \au{Bae, Hyunji~Jane}} \yr{2019}  \at{Characteristic scales of townsend’s wall-attached eddies}.  \jt{Journal of fluid mechanics}  \bvol{868},  \pg{698--725}.

\bibitem[Marcus(1993)]{marcus1993jupiter}
{\sc \au{Marcus, Philip~S}} \yr{1993}  \at{Jupiter's great red spot and other vortices}.  \jt{Annual Review of Astronomy and Astrophysics}  \bvol{31}~(1),  \pg{523--569}.

\bibitem[Motoori \& Goto(2019)]{motoori2019generation}
{\sc \au{Motoori, Yutaro} \& \au{Goto, Susumu}} \yr{2019}  \at{Generation mechanism of a hierarchy of vortices in a turbulent boundary layer}.  \jt{Journal of Fluid Mechanics}  \bvol{865},  \pg{1085--1109}.

\bibitem[Neamtu-Halic {\em et~al.\/}(2019)Neamtu-Halic, Krug, Haller \& Holzner]{halic_krug_haller_holzner_2019}
{\sc \au{Neamtu-Halic, Marius~M.}, \au{Krug, Dominik}, \au{Haller, George} \& \au{Holzner, Markus}} \yr{2019}  \at{Lagrangian coherent structures and entrainment near the turbulent/non-turbulent interface of a gravity current}.  \jt{Journal of Fluid Mechanics}  \bvol{877},  \pg{824–843}.

\bibitem[Onu {\em et~al.\/}(2015)Onu, Huhn \& Haller]{ONU201526}
{\sc \au{Onu, K.}, \au{Huhn, F.} \& \au{Haller, G.}} \yr{2015}  \at{{LCS} tool: A computational platform for lagrangian coherent structures}.  \jt{Journal of Computational Science}  \bvol{7},  \pg{26--36}.

\bibitem[Pan {\em et~al.\/}(2009)Pan, Wang \& Zhang]{pan2009identification}
{\sc \au{Pan, Chong}, \au{Wang, JinJun} \& \au{Zhang, Cao}} \yr{2009}  \at{Identification of lagrangian coherent structures in the turbulent boundary layer}.  \jt{Science in China Series G: Physics, Mechanics and Astronomy}  \bvol{52}~(2),  \pg{248--257}.

\bibitem[Park(2017)]{park2017wall}
{\sc \au{Park, George~Ilhwan}} \yr{2017}  \at{Wall-modeled large-eddy simulation of a high reynolds number separating and reattaching flow}.  \jt{AIAA Journal}  \bvol{55}~(11),  \pg{3709--3721}.

\bibitem[Park \& Moin(2014)]{park2014improved}
{\sc \au{Park, George~Ilhwan} \& \au{Moin, Parviz}} \yr{2014}  \at{An improved dynamic non-equilibrium wall-model for large eddy simulation}.  \jt{Physics of Fluids}  \bvol{26}~(1).

\bibitem[Park \& Moin(2016)]{Park2016}
{\sc \au{Park, G.~I.} \& \au{Moin, P.}} \yr{2016}  \at{Numerical aspects and implementation of a two-layer zonal wall model for les of compressible turbulent flows on unstructured meshes}.  \jt{\emph{J. Comput. Phys.}}  \bvol{305},  \pg{589--603}.

\bibitem[Pierce {\em et~al.\/}(2013)Pierce, Moin \& Sayadi]{pierce2013application}
{\sc \au{Pierce, Brian}, \au{Moin, Parviz} \& \au{Sayadi, Taraneh}} \yr{2013}  \at{Application of vortex identification schemes to direct numerical simulation data of a transitional boundary layer}.  \jt{Physics of Fluids}  \bvol{25}~(1),  \pg{015102}.

\bibitem[Ran {\em et~al.\/}(2021)Ran, Brosseau, Blackwell, Qin, Winter \& Arratia]{ran2021bacteria}
{\sc \au{Ran, Ranjiangshang}, \au{Brosseau, Quentin}, \au{Blackwell, Brendan~C}, \au{Qin, Boyang}, \au{Winter, Rebecca~L} \& \au{Arratia, Paulo~E}} \yr{2021}  \at{Bacteria hinder large-scale transport and enhance small-scale mixing in time-periodic flows}.  \jt{Proceedings of the National Academy of Sciences}  \bvol{118}~(40),  \pg{e2108548118}.

\bibitem[Rockwood {\em et~al.\/}(2018)Rockwood, Huang \& Green]{rockwood2018tracking}
{\sc \au{Rockwood, Matthew}, \au{Huang, Yangzi} \& \au{Green, Melissa}} \yr{2018}  \at{Tracking coherent structures in massively-separated and turbulent flows}.  \jt{Physical Review Fluids}  \bvol{3}~(1),  \pg{014702}.

\bibitem[Rycroft {\em et~al.\/}(2020)Rycroft, Wu, Yu \& Kamrin]{rycroft_reference_2020}
{\sc \au{Rycroft, Chris~H.}, \au{Wu, Chen-Hung}, \au{Yu, Yue} \& \au{Kamrin, Ken}} \yr{2020}  \at{Reference map technique for incompressible fluid–structure interaction}.  \jt{J. Fluid Mech.}  \bvol{898},  \pg{A9}.

\bibitem[Schl{\"u}ter {\em et~al.\/}(2005)Schl{\"u}ter, Pitsch \& Moin]{schluter2005outflow}
{\sc \au{Schl{\"u}ter, JU}, \au{Pitsch, H} \& \au{Moin, P}} \yr{2005}  \at{Outflow conditions for intregrated large eddy simulation/reynolds-averaged navier-stokes simulations}.  \jt{AIAA journal}  \bvol{43}~(1),  \pg{156--164}.

\bibitem[Shadden {\em et~al.\/}(2005)Shadden, Lekien \& Marsden]{shadden_definition_2005}
{\sc \au{Shadden, Shawn~C.}, \au{Lekien, Francois} \& \au{Marsden, Jerrold~E.}} \yr{2005}  \at{Definition and properties of {Lagrangian} coherent structures from finite-time {Lyapunov} exponents in two-dimensional aperiodic flows}.  \jt{Physica D: Nonlinear Phenomena}  \bvol{212}~(3-4),  \pg{271--304}.

\bibitem[Sousa {\em et~al.\/}(2016)Sousa, Poole, Afonso, Pinho, Oliveira, Morozov \& Alves]{sousa2016lid}
{\sc \au{Sousa, RG}, \au{Poole, RJ}, \au{Afonso, AM}, \au{Pinho, FT}, \au{Oliveira, PJ}, \au{Morozov, A} \& \au{Alves, MA}} \yr{2016}  \at{Lid-driven cavity flow of viscoelastic liquids}.  \jt{Journal of Non-Newtonian Fluid Mechanics}  \bvol{234},  \pg{129--138}.

\bibitem[Thomas \& David(2022)]{thomas2022eulerian}
{\sc \au{Thomas, L} \& \au{David, L}} \yr{2022}  \at{Eulerian and lagrangian coherent structures in a positive surge}.  \jt{Experiments in Fluids}  \bvol{63}~(2),  \pg{43}.

\bibitem[Urzay {\em et~al.\/}(2017)Urzay, Doostmohammadi \& Yeomans]{urzay2017multi}
{\sc \au{Urzay, Javier}, \au{Doostmohammadi, Amin} \& \au{Yeomans, Julia~M}} \yr{2017}  \at{Multi-scale statistics of turbulence motorized by active matter}.  \jt{Journal of Fluid Mechanics}  \bvol{822},  \pg{762--773}.

\bibitem[Valkov {\em et~al.\/}(2015)Valkov, Rycroft \& Kamrin]{valkov_eulerian_2015}
{\sc \au{Valkov, Boris}, \au{Rycroft, Chris~H.} \& \au{Kamrin, Ken}} \yr{2015}  \at{Eulerian {Method} for {Multiphase} {Interactions} of {Soft} {Solid} {Bodies} in {Fluids}}.  \jt{Journal of Applied Mechanics}  \bvol{82}~(4),  \pg{041011}.

\bibitem[Wang {\em et~al.\/}(2022)Wang, Kamrin \& Rycroft]{wang2022incompressible}
{\sc \au{Wang, Xiaolin}, \au{Kamrin, Ken} \& \au{Rycroft, Chris~H}} \yr{2022}  \at{An incompressible eulerian method for fluid--structure interaction with mixed soft and rigid solids}.  \jt{Physics of Fluids}  \bvol{34}~(3).

\bibitem[Wilson {\em et~al.\/}(2013)Wilson, Tutkun \& Cal]{wilson_tutkun_cal_2013}
{\sc \au{Wilson, Z.~D.}, \au{Tutkun, M.} \& \au{Cal, R.~B.}} \yr{2013}  \at{Identification of lagrangian coherent structures in a turbulent boundary layer}.  \jt{Journal of Fluid Mechanics}  \bvol{728},  \pg{396–416}.

\bibitem[Yang {\em et~al.\/}(2010)Yang, Pullin \& Bermejo-Moreno]{yang2010multi}
{\sc \au{Yang, Yue}, \au{Pullin, DI} \& \au{Bermejo-Moreno, Ivan}} \yr{2010}  \at{Multi-scale geometric analysis of lagrangian structures in isotropic turbulence}.  \jt{Journal of Fluid Mechanics}  \bvol{654},  \pg{233--270}.

\bibitem[Yao \& Hussain(2020)]{yao2020physical}
{\sc \au{Yao, Jie} \& \au{Hussain, Fazle}} \yr{2020}  \at{A physical model of turbulence cascade via vortex reconnection sequence and avalanche}.  \jt{Journal of Fluid Mechanics}  \bvol{883},  \pg{A51}.

\bibitem[Yeung(2002)]{yeung2002lagrangian}
{\sc \au{Yeung, PK}} \yr{2002}  \at{Lagrangian investigations of turbulence}.  \jt{Annual review of fluid mechanics}  \bvol{34}~(1),  \pg{115--142}.

\bibitem[You \& Leung(2014)]{you2014eulerian}
{\sc \au{You, Guoqiao} \& \au{Leung, Shingyu}} \yr{2014}  \at{An eulerian method for computing the coherent ergodic partition of continuous dynamical systems}.  \jt{Journal of Computational Physics}  \bvol{264},  \pg{112--132}.

\bibitem[You \& Leung(2018{\natexlab{{\em a\/}}})]{you2018eulerian}
{\sc \au{You, Guoqiao} \& \au{Leung, Shingyu}} \yr{2018{\natexlab{{\em a\/}}}}  \at{Eulerian based interpolation schemes for flow map construction and line integral computation with applications to lagrangian coherent structures extraction}.  \jt{Journal of Scientific Computing}  \bvol{74},  \pg{70--96}.

\bibitem[You \& Leung(2018{\natexlab{{\em b\/}}})]{you2018improved}
{\sc \au{You, Guoqiao} \& \au{Leung, Shingyu}} \yr{2018{\natexlab{{\em b\/}}}}  \at{An improved eulerian approach for the finite time lyapunov exponent}.  \jt{Journal of Scientific Computing}  \bvol{76}~(3),  \pg{1407--1435}.

\bibitem[You \& Leung(2020)]{you2020fast}
{\sc \au{You, Guoqiao} \& \au{Leung, Shingyu}} \yr{2020}  \at{Fast construction of forward flow maps using eulerian based interpolation schemes}.  \jt{Journal of Scientific Computing}  \bvol{82}~(2),  \pg{32}.

\bibitem[You \& Leung(2021)]{you2021eulerian}
{\sc \au{You, Guoqiao} \& \au{Leung, Shingyu}} \yr{2021}  \at{Eulerian algorithms for computing some lagrangian flow network quantities}.  \jt{Journal of Computational Physics}  \bvol{445},  \pg{110620}.

\bibitem[Zhou {\em et~al.\/}(1999)Zhou, Adrian, Balachandar \& Kendall]{zhou1999mechanisms}
{\sc \au{Zhou, Jigen}, \au{Adrian, Ronald~J}, \au{Balachandar, Sivaramakrishnan} \& \au{Kendall, TM1693393}} \yr{1999}  \at{Mechanisms for generating coherent packets of hairpin vortices in channel flow}.  \jt{Journal of fluid mechanics}  \bvol{387},  \pg{353--396}.

\end{thebibliography}

\end{document}